\documentclass[usenatbib,fleqn,useAMS]{mn2e}

\usepackage{amsmath}
\usepackage[varg]{txfonts}
\usepackage{mathrsfs}
\usepackage{braket}

\usepackage[OT2,OT1]{fontenc}
\newcommand\cyr{
\renewcommand\rmdefault{wncyr}
\renewcommand\sfdefault{wncyss}
\renewcommand\encodingdefault{OT2}
\normalfont
\selectfont}
\DeclareTextFontCommand{\textcyr}{\cyr}

\renewcommand\pi\upi
\renewcommand\partial\upartial
\newcommand\Nu{\tilde\nu}
\newcommand\rmd{\mathrm d}
\newcommand\rme{\mathrm e}

\providecommand\flr[1]{\lfloor{#1}\rfloor}
\providecommand\clg[1]{\lceil{#1}\rceil}

\newtheorem{definition}{Definition}[section]
\newtheorem{theorem}[definition]{Theorem}
\newtheorem{corollary}[definition]{Corollary}
\newtheorem{lemma}[definition]{Lemma}

\title[phase-space consistency of augmented densities]
{Phase-space consistency of stellar dynamical models
determined by separable augmented densities}

\author[An, Van~Hese, \& Baes]
{J.~An$^1$\thanks{E-mail:~jinan@nao.cas.cn}, E.~Van~Hese$^2$, 
and M.~Baes$^2$
\\$^1$ National Astronomical Observatories,
Chinese Academy of Sciences, A20 Datun Road, Chaoyang District,
Beijing 100012, PR~China;
\\$^2$ Sterrenkundig Observatorium, Universiteit Gent (Ghent University),
Kr{\ij}gslaan 281 S9, Gent (Ghent),
B-9000, Belgium}

\pagerange{\pageref{firstpage}--\pageref{lastpage}}
\pubyear{2012}

\begin{document}
\label{firstpage}
\maketitle

\begin{abstract}
Assuming the separable augmented density, it is always possible
to construct a distribution function of a spherical population
with any given density and anisotropy. We consider under
what conditions the distribution constructed as such is in fact
non-negative everywhere in the accessible phase-space.
We first generalize known necessary conditions
on the augmented density using fractional calculus.
The condition on the radius part $R(r^2)$ (whose logarithmic
derivative is the anisotropy parameter) is equivalent
to the complete monotonicity of $w^{-1}R(w^{-1})$.
The condition on the potential part on the other hand is given by
its derivative up to any order not greater than $\frac32-\beta_0$
being non-negative where $\beta_0$ is the central anisotropy parameter.
We also derive a specialized inversion formula for the distribution
from the separable augmented density, which leads to sufficient
conditions on separable augmented densities for the non-negativity
of the distribution. The last generalizes the similar condition derived
earlier for the generalized Cuddeford system to arbitrary separable systems.
\end{abstract}

\section{Introduction}

Except maybe in our imagination is nothing in our universe exactly
spherically symmetric. Yet spherical models by virtue of simplicity
have widely been adopted as the default route when we embark on
something new to investigate. What is surprising is that insights
obtained from these `spherical cows' appear to be
helpful at all for our understanding of the `real' universe.
This is particularly true for dynamical models of stellar systems.
Models of spherical stellar systems are
not only useful to approximate putative dark haloes or any actual
roundish aggregate system found in the sky but also important
to provide the simplest test ground for the physical principles and
understanding of structures governed by them.

It was \citet{De86} who had first used augmented densities
(i.e., extensions of the density profile into bivariate functions
of the potential and radius) of a spherical system to build
a dynamical model of spherical stellar systems.
Whilst the information contained in the distribution function and
the corresponding augmented density is mathematically equivalent,
the approach through the augmented density, in particular for
such systems with anisotropic velocity distributions, is advantageous
since its relations to directly observable quantities are
simpler than those of the distribution function. That is to say,
it is in principle trivial to find an augmented density with desired
behaviours of observables unlike distribution functions, observables
resulting from which are only available through moment integrals.
For example, \emph{an} augmented density $\Nu(\Psi,r^2)$
(and subsequently a distribution function via algorithmic inversions)
can be found from arbitrarily specified profiles of the density $\nu(r)$
and the anisotropy parameter such that $\Nu(\Psi,r^2)=P(\Psi)R(r^2)$
where $P[\Psi(r)]=\nu(r)/R(r^2)$ and $R(r^2)$ is given by
equation (\ref{eq:beta}) from the prescribed anisotropy
\citep{QH95,BvH07}.

A drawback of this approach is that one does
not know a priori whether the spherical system described by
the given augmented density is consistent with being built by
a physical distribution, that is, non-negative everywhere
in the accessible phase space (\emph{the phase-space consistency}).
For some systems however where the inversion algorithm reduces
to a single integral quadrature such as the constant anisotropy system
\citep[see e.g.,][]{EA06}, the criteria on the augmented
density for the phase-space consistency have been derived.
For instance, \citet{CP92} had discovered necessary and sufficient
conditions for the non-negativity of the
\citeauthor{Os79}--\citeauthor{Me85} distribution function
expressed in terms of the corresponding augmented density,
and \citet{CM10} extended these to be applicable to
the multicomponent generalized \citeauthor{Cu91} system.
\citet{CM10b} have essentially hypothesized that 
the necessary conditions of \citet{CM10}, which concerns the
behaviour of the potential-dependent parts of augmented densities,
may be applicable to any system for which the potential and radial
dependencies of the augmented density are multiplicatively separable.
This has been subsequently proven by \citet{vHBD11} and \citet{An11a}
whereas \citet{An11} was able to find necessary conditions
on the radius dependent parts of separable augmented densities,
which results in the constraints on the behaviour of the anisotropy
parameter that can be consistent with separable augmented densities.

This paper continues the study of the phase-space consistency
criteria for separable augmented densities. As its logical conclusion,
we attempt to provide an answer to the question,
under what conditions the distribution function constructed from a
separable augmented density is non-negative everywhere in the entire
accessible subvolume of the phase space. This paper is organized as follows.
We start by reviewing the concepts of the distribution function
and the augmented density in Sect.~\ref{sec:mod}, in which
we also present a result (eq.~\ref{eq:dpr}) that leads to many of
main arguments. Using this, first in Sect.~\ref{sec:moms} we elucidate
the relation amongst the distribution function, the augmented density,
and the observables. The main findings of this paper are provided
in Sect.~\ref{sec:nec} where necessary conditions on separable
augmented densities for the phase-space consistency are presented,
and in Sect.~\ref{sec:suf} where corresponding sufficient conditions
are given. In Sect.~\ref{sec:rfun} we present an application
on a parameterization of the anisotropy suitable for
practical modelling. This paper concludes with the summary
of findings in Sect.~\ref{sec:sum}.
Mathematical ideas used in this paper reviewed in Appendices.

\section{Models for spherical dynamical systems}
\label{sec:mod}
\subsection{Distribution function}

Let $F(\bmath r;\bmath v)$
be a steady-state phase-space distribution such that
$\int_SF\,\rmd^3\!\bmath r\,\rmd^3\!\bmath v$
is the number of tracers in any measurable phase-space volume $S$.
Here $\bmath r$ is the position vector in the configuration space and
$\bmath v=\dot{\bmath r}$ is the velocity. Assuming spherical symmetry,
the distribution is invariant under any orthogonal transformation,
which implies that
$F(\bmath r;\bmath v)=F(r;v_r,v_\mathrm t)$
where $r=\lVert{\bmath r}\rVert$ is the radial distance,
$v_r=\bmath{v\cdot}\hat{\bmath r}$ and
$v_\mathrm t=\lVert{\bmath v-v_r\hat{\bmath r}}\rVert$ are the radial and
tangential velocities with $\hat{\bmath r}=\bmath r/r$ being the radial
unit vector. If we adopt the spherical polar coordinate
$(r,\theta,\phi)$, these are also given by
$\lVert{\bmath v}\rVert^2=v^2=v_r^2+v_\mathrm t^2$ and
$v_\mathrm t^2=v_\theta^2+v_\phi^2$
where $(v_r,v_\theta,v_\phi)=(\dot r,r\dot\theta,r\dot\phi\sin\theta)$
are the velocity components projected onto the associated orthonormal basis.
Moreover, the \citeauthor{Je15} theorem indicates that if the given
distribution function (df) is a solution to the collisionless Boltzmann
equation with a \emph{generic} static spherical potential $\Phi(r)$,
it must be in the form of $F(E,L^2)$
where $E=\Psi(r)-v^2/2$ and $L=rv_\mathrm t$
are the two isotropic isolating integrals admitted by all generic
static spherical potentials, namely the specific binding energy
and the magnitude of the specific angular momentum. Here,
\begin{equation}
\Psi(r)\equiv\begin{cases}
\Phi(r_\mathrm{out})-\Phi(r)&\text{if $r_\mathrm{out}$ is finite}\\
\Phi(\infty)-\Phi(r)&
\text{if $r_\mathrm{out}=\infty$ and $\lvert{\Phi(\infty)}\rvert<\infty$}\\
-\Phi(r)&\text{if $r_\mathrm{out}=\infty$ and
$\Phi(\infty)\rightarrow\infty$}\\
\end{cases}
\end{equation}
is the relative potential with respect to the boundary $r_\mathrm{out}$.
The system that is not confined within a finite boundary radius
is represented by $r_\mathrm{out}=\infty$ with
$\Phi(\infty)=\lim_{r\rightarrow\infty}\Phi(r)$.
If $r_\mathrm{out}$ or $\Phi(\infty)$ is finite, then
$F(E<0,L^2)=0$ because by definition $E\ge0$
for all tracers bound to the system (and
bounded by $r\le r_\mathrm{out}$).

\subsection{Augmented density}

Integrating $F(E,L^2)$ over the velocity space results in
\begin{equation}\label{eq:ad}
\Nu(\Psi,r^2)\equiv\iiint\!\rmd^3\!\bmath v\,
F\bigl(E=\Psi-\tfrac12v^2,L^2=r^2v_\mathrm t^2\bigr),
\end{equation}
a bivariate function of $\Psi$ and $r^2$,
that is, the \emph{augmented density} (AD).
The integral is over the whole velocity
subspace, but if $r_\mathrm{out}$ or $\Phi(\infty)$ is finite, it is
essentially within the sphere $v^2\le2\Psi$ since
$F(E<0,L^2)=0$ for these cases. With $\Psi(r)$ specified,
the AD yields the local density via $\nu(r)=\Nu[\Psi(r),r^2]$.
Similarly, the augmented moment functions
(n.b., $\Nu=m_{0,0}$) are given by
\begin{subequations}
\begin{multline}
m_{k,n}(\Psi,r^2)\equiv\iiint\!\rmd^3\!\bmath v\,v_r^{2k}v_\mathrm t^{2n}
F\bigl(E=\Psi-\tfrac12v^2,L^2=r^2v_\mathrm t^2\bigr)
\\=4\pi\!\iint_{v_r\ge0,v_\mathrm t\ge0,(v^2\le2\Psi)}\!
\rmd v_r\,\rmd v_\mathrm t\,v_r^{2k}v_\mathrm t^{2n+1}
F\Bigl(\Psi-\frac{v_r^2+v_\mathrm t^2}2,r^2v_\mathrm t^2\Bigr).
\end{multline}
Changing the integration variables to $(E,L^2)$,
these are represented to be a set of integral transformations of the df,
\begin{equation}\label{eq:dist}\begin{split}
m_{k,n}&=\frac{2\pi}{r^{2n+2}}\!
\iint_T\!\rmd E\,\rmd L^2
K^{k-\frac12}L^{2n}F(E,L^2)
\\&=\frac{2\pi}{r^{2n+2}}\!
\iint_{E\ge E_0,L^2\ge0}\!
\rmd E\,\rmd L^2\Theta(K)\,
\lvert{K}\rvert^{k-\frac12}L^{2n}F(E,L^2).
\end{split}\end{equation}
\end{subequations}
Here $\Theta(x)$ is the Heaviside unit-step function and
\begin{equation}
E_0\equiv\begin{cases}
0&\text{if $r_\mathrm{out}$ or $\Phi(\infty)$ is finite}\\
-\infty&\text{if}\
\lim_{r\rightarrow\infty}\Psi(r)=-\Phi(\infty)\rightarrow-\infty
\end{cases}
\end{equation}
is the lower bound of the binding energy.
The transform kernel is $K(E,L^2;\Psi,r^2)\equiv
2(\Psi-E)-L^2r^{-2}$,
which is $v_r^2$ expressed as a function of $4$-tuple
$(E,L^2;\Psi,r^2)$. Finally, the domain of $(E,L^2)$
space in which the integral is performed is
$T\equiv\set{(E,L^2)|E\ge E_0,L^2\ge0,K\ge0}$.


\citet{An11a} has shown that the Abel transformation of the
augmented moment function results in an integral transformation of
the df similar to equation (\ref{eq:dist}) but with
different powers on $K$ and $L^2$.
This is generalized by means of \emph{fractional calculus}
(Appendix \ref{app:frd}), that is,
for any pair of non-negative reals $\xi\ge\mu\ge0$,
\begin{subequations}\label{eq:dpr}
\begin{gather}\label{eq:dpirn2}
{{}_{E_0}D_\Psi}^\mu\biggl[
{{}_0I_{r^2}}^{\xi-\frac12}\Bigl(\frac{\Nu}{r^{2\xi-1}}\Bigr)\biggr]
\\\nonumber\quad=\begin{cases}{\displaystyle
\frac{2^{\mu+1}\pi^\frac32r^{2\xi-3}}{\Gamma(\xi-\mu)}\!
\iint_T\!\rmd E\,\rmd L^2
\frac{K^{\xi-\mu-1}}{L^{2\xi-1}}F(E,L^2)
}&(\xi>\mu)\\{\displaystyle
2^\xi\pi^\frac32r^{2\xi-3}\!
\int_0^{L_\mathrm m^2}\!\frac{\rmd L^2}{L^{2\xi-1}}
F\Bigl(\Psi-\frac{L^2}{2r^2},L^2\Bigr)
}&(\xi=\mu)\end{cases},
\\\label{eq:dripn2}
{{}_0D_{r^2}}^\mu\Bigl(
r^{2\mu}{{}_{E_0}I_\Psi}^{\xi-\frac12}\Nu\Bigr)
\\\nonumber\quad=\begin{cases}{\displaystyle
\frac{2^{\frac32-\xi}\pi^\frac32}{r^{2\mu+2}\Gamma(\xi-\mu)}\!
\iint_T\!\rmd E\,\rmd L^2
K^{\xi-\mu-1}L^{2\mu}F(E,L^2)
}&(\xi>\mu)\\{\displaystyle
\frac{\pi^\frac32}{2^{\mu-\frac12}r^{2\mu+2}}\!
\int_0^{L_\mathrm m^2}\!\rmd L^2L^{2\mu}
F\Bigl(\Psi-\frac{L^2}{2r^2},L^2\Bigr)
}&(\xi=\mu)\end{cases}
\end{gather}
\end{subequations}
where $\Gamma(x)$ is the gamma function and
the operators ${{}_aI_x}^\lambda$ and ${{}_aD_x}^\lambda$ are
as defined in Appendix \ref{app:frd}. In addition,
\begin{equation}
L_\mathrm m^2\equiv\begin{cases}
2r^2\Psi&\text{if $E_0=0$}\\
\infty&\text{if $E_0=-\infty$}\end{cases}.
\end{equation}
Derivations are provided in Appendix \ref{sec:cal}.

\section{Moment sequences \& augmented densities}
\label{sec:moms}

The knowledge of $\Nu(\Psi,r^2)$ is mathematically
equivalent to knowing $F(E,L^2)$.
In particular, once the potential $\Psi=\Psi(r)$ is specified,
the specification of the AD completely determine
a unique spherical dynamic system in equilibrium.
In light of equation (\ref{eq:dpr}), here we seek
a possible `physical interpretation' of the AD in relation to the df
for describing dynamic systems.

Consider the moment sequence of the df restricted along $K=0$,
\begin{subequations}\label{eq:moms}
\begin{multline}
\mathscr M_\mu(\Psi,r^2)\equiv\frac{(2\pi)^\frac32}{(2r^2)^{\mu+1}}\!
\int_0^{L_\mathrm m^2}\!\rmd L^2
L^{2\mu}F\Bigl(\Psi-\frac{L^2}{2r^2},\,L^2\Bigr)
\\=\begin{cases}{\displaystyle
\Psi^{\mu+1}\!\int_0^1\!\rmd y\,y^\mu\mathscr F(y\Psi;\Psi,r^2)
}&(E_0=0,\,L_\mathrm m^2=2r^2\Psi)\\{\displaystyle
\int_0^\infty\!\rmd Y\,Y^\mu\mathscr F(Y;\Psi,r^2)
}&(E_0=-\infty,\,L_\mathrm m^2=\infty)
\end{cases},
\end{multline}
where
\begin{equation}
\mathscr F(Y;\Psi,r^2)\equiv
(2\pi)^\frac32F(\Psi-Y,2r^2Y).
\end{equation}
\end{subequations}
Then equations (\ref{eq:dpr}) indicate that
\begin{subequations}\label{eq:msad}
\begin{equation}
\mathscr M_\mu=\begin{cases}
{{}_{E_0}I_\Psi}^{\mu-1/2}
{{}_0D_{r^2}}^\mu\bigl(r^{2\mu}\Nu\bigr)
&(\mu\ge\frac12)\\
{{}_{E_0}D_\Psi}^{1/2-\mu}
{{}_0D_{r^2}}^\mu\bigr(r^{2\mu}\Nu\bigr)
&(0\le\mu\le\frac12)\\
{{}_{E_0}D_\Psi}^{\xi+1/2}
{{}_0I_{r^2}}^\xi
\Bigl(\dfrac{\Nu}{r^{2\xi}}\Bigr)
&(\xi=-\mu\ge0)\end{cases}.
\end{equation}
In particular, if $\mu$ is a non-negative integer,
this results in
\begin{equation}\begin{split}
\mathscr M_0&=\frac1{\sqrt\pi}\frac\partial{\partial\Psi}\!
\int_{E_0}^\Psi\!
\frac{\Nu(Q,r^2)\,\rmd Q}{\sqrt{\Psi-Q}}
\\
\mathscr M_n
&=\frac1{\bigl(\tfrac12\bigr)_{n-1}^+\!\sqrt\pi}
\int_{E_0}^\Psi\!\rmd Q\,
(\Psi-Q)^{n-\frac32}
\biggl(\frac\partial{\partial r^2}\biggr)^n\bigl[r^{2n}\Nu(Q,r^2)\bigr],
\end{split}\end{equation}
\end{subequations}
where $n=1,2,\dotsc$ and $(a)^+_n=\prod_{j=1}^n(a-1+j)$
is the \emph{rising} sequential product. In other words,
$\Nu(\Psi,r^2)$ directly determine the entire moment sequences
along a fixed sectional line in $(E,L^2)$ space.
The AD in this sense is similar to the \emph{moment generating function}
or the \emph{characteristic function} for the df as a probability density.
With varying $(\Psi,r^2)$, the $K=0$ lines
sweep the whole accessible $(E,L^2)$ space,
and thus $\Nu(\Psi,r^2)$ in principle uniquely determines $F(E,L^2)$.
Explicit inversion algorithms from $\Nu(\Psi,r^2)$ to $F(E,L^2)$
are available in literature utilizing either the known inverse
of named integral transforms \citep[e.g.,][]{Ly62,De86}
or complex contour integrals \citep[e.g.,][]{HQ93}.

Next, we consider what information on physical properties
of the system is sufficient to specify a unique AD. For this,
equation (\ref{eq:dripn2}) indicates that
the even-order (augmented) velocity moments are related to
the AD as in \citep[eq.~13]{DM92}
\begin{subequations}
\begin{equation}\label{eq:vmts}\begin{split}
m_{k,n}(\Psi,r^2)&
=\frac{2^{k+n}\Gamma(k+\frac12)}{\sqrt\pi r^{2n+2}}
\biggl(r^4\!\frac\partial{\partial r^2}\biggr)^n
\bigl(r^2{{}_{E_0}I_\Psi}^{n+k}\Nu\bigr)
\\&=2^{k+n}\bigl(\tfrac12\bigr)_k^+
{{}_{E_0}I_\Psi}^{k+n}
\bigl[{{}_0D_{r^2}}^n(r^{2n}\Nu)\bigr],
\end{split}\end{equation}
Here note $\sqrt\pi\,(\frac12)_k^+=\Gamma(k+\frac12)$.
Given the potential $\Psi(r)$,
specifying the AD completely fixes every (in principle observable)
velocity moment with equation (\ref{eq:vmts}) such that
\begin{equation}
\overline{v_r^{2k}v_\mathrm t^{2n}}=
\frac{m_{k,n}[\Psi(r),r^2]}{\Nu[\Psi(r),r^2]}.
\end{equation}
\end{subequations}
Conversely, equation (\ref{eq:vmts}) for $(k,n)=(\mu+1,0)$, that is,
$m_{\mu+1,0}=2^{\mu+1}(\frac12)_{\mu+1}^+
{{}_{E_0}I_\Psi}^{\mu+1}\Nu$ \emph{at a fixed $r$} reduces to
\begin{subequations}
\begin{equation}\begin{split}
\mathscr V_\mu(r)&\equiv
\frac{\mu!\overline{v_r^{2(\mu+1)}}}
{2^{\mu+1}\bigl(\frac12\bigr)_{\mu+1}^+}
\\&=\begin{cases}{\displaystyle
[\Psi(r)]^{\mu+1}\!\int_0^1\!\rmd q\,q^\mu\mathscr P\bigl[q\Psi(r);r\bigr]
}&(E_0=0)\\{\displaystyle
\int_0^\infty\!\rmd Q\,Q^\mu\mathscr P(Q;r)
}&(E_0=-\infty)
\end{cases},\end{split}
\end{equation}
where
\begin{equation}
\mathscr P(Q;r)\equiv\frac{\Nu[\Psi(r)-Q,r^2]}{\nu(r)}.
\end{equation}
\end{subequations}
%
That is, given the local density $\nu(r)$ and the potential $\Psi(r)$,
the infinite set of the radial velocity moments
in every order consists in the moment sequence of
the AD considered as a distribution of $\Psi$
at fixed $r$. The problem is reducible to the \emph{Hausdorff} 
(for $E_0=0$) or the \emph{Stieltjes {\rm(for $E_0=-\infty$)} moment problems}.
With the infinite sequence of the radial velocity moments
as functions of $r$, the AD can then be
uniquely determined at least formally by such means as
e.g., the Hilbert basis or the Laplace and/or Fourier transform (cf.,
the moment generating function and the characteristic function) etc.

The final information required for the full specification of the system
is the determination of the potential.
The \emph{self-consistent} potential may be determined through
the Poisson equation: that is, if the mass-to-light ratio is constant,
$\Psi(r)$ can be fixed by solving the ordinary differential equation
on $\Psi(r)$ that results from the spherical Poisson equation with
the source term given by $\nu=\Nu(\Psi,r^2)$. Alternatively, from
equation (\ref{eq:vmts}), we deduce for $k\ge1$ that
\begin{subequations}
\begin{gather}
\frac{\partial m_{k,n}}{\partial\Psi}
=(2k-1)\,m_{k-1,n};
\nonumber\\
\frac{\partial(r^{2n+2}m_{k,n})}{\partial r^2}
=\bigl(k-\tfrac12\bigr)\,r^{2n}m_{k-1,n+1}.
\end{gather}
Consequently the total radial derivative of $m_{k,n}$ for $k\ge1$
results in
\begin{multline}
\frac{\rmd m_{k,n}}{\rmd r}
=\frac{2m_{k,n}}{r}\biggl[
\frac{\partial\log(r^{2n+2}m_{k,n})}{\partial\log r^2}
-(n+1)\biggr]
+\frac{\rmd\Psi}{\rmd r}\frac{\partial m_{k,n}}{\partial\Psi}
\\=-\frac{2(n+1)m_{k,n}-(2k-1)m_{k-1,n+1}}r
+(2k-1)m_{k-1,n}\frac{\rmd\Psi}{\rmd r}.
\end{multline}
\end{subequations}
With $\Psi=\Psi(r)$ and
$m_{k,n}[\Psi(r),r^2]=\nu\overline{v_r^{2k}v_\mathrm t^{2n}}$,
this may be solved for $\rmd\Psi/\rmd r$ if the required velocity
moments as a function of $r$ are known. For the simplest case
$(k,n)=(1,0)$, this reduces to the spherical
(second-order steady-state) Jeans equation.
%
%

\section{Necessary conditions for separable augmented densities}
\label{sec:nec}

In the following, we limit our concern to the cases for which
the potential and the radius dependencies of the AD are
multiplicatively separable such that
\begin{equation}\label{eq:sep}
\Nu(\Psi,r^2)=P(\Psi)R(r^2).
\end{equation}
In addition to mathematical expediency, this assumption is also notable
because under the separability assumption in equation (\ref{eq:sep}),
the radius part $R(r^2)$ of the AD alone uniquely specifies
the so-called \citeauthor{Bi80} anisotropy parameter,
\begin{gather}\begin{split}
\beta(r)\equiv1-\frac{\overline{v_\mathrm t^2}}{2\overline{v_r^2}}
&=1-\frac{m_{0,1}[\Psi(r),r^2]}{2m_{1,0}[\Psi(r),r^2]}
\\&=1-\frac1{m_{1,0}}\frac{\partial(r^2m_{1,0})}{\partial r^2}
=-\frac{\partial\log m_{1,0}}{\partial\log r^2}\biggr\rvert_{\Psi(r),r^2}
\end{split}
\intertext{such that \citep{De86,QH95}}
\beta(r)=-\frac{\rmd\log R(r^2)}{\rmd\log r^2}
\,;\qquad
\frac{R(r^2)}{R(r_0^2)}
=\exp\biggl\lgroup\int_r^{r_0}\!\frac{2\beta(s)}s\rmd s\biggr\rgroup.
\label{eq:beta}
\end{gather}
Some applications are found in \citet{BvH07}
whilst \citet{An11} discusses implications of the separability
assumption.

\subsection{Conditions on the radius part}
\label{sec:radn}

\citet{An11} has argued that (hereafter $x\equiv r^2$)
\begin{equation}\label{eq:main1}
R_{(n)}(x)\equiv\frac{\rmd^n[x^nR(x)]}{\rmd x^n}\ge0
\qquad(x>0,\,n=0,1,2,\dotsc)
\end{equation}
for the radius part $R(x)$ of equation (\ref{eq:sep})
is necessary for the non-negativity of the corresponding df.
Here we derive several equivalent statements of this condition.

First of these is
\begin{equation}\label{eq:main0}
{{}_0D_x}^\mu(x^\mu R)\ge0
\qquad(x>0,\,\mu\ge0).
\end{equation}
This follows equation (\ref{eq:dripn2}),
which indicates that for $0\le\mu\le\xi$
\begin{equation}
{{}_0D_x}^\mu\bigl(
x^\mu{{}_{E_0}I_\Psi}^{\xi-\frac12}\Nu\bigr)
={{}_{E_0}I_\Psi}^{\xi-\frac12}P(\Psi)\cdot
{{}_0D_x}^\mu[x^\mu R(x)]\ge0
\end{equation}
given equation (\ref{eq:sep}). Since $P\ge0$ is obviously necessary,
equation (\ref{eq:main0}) follows this and Lemma \ref{lem:pos},
which implies that ${{}_{E_0}I_\Psi}^{\xi-\frac12}P>0$ for $\xi\ge\frac12$.
It is trivial that equation (\ref{eq:main0}) implies
equation (\ref{eq:main1}) as the latter is the restriction of
the former for an integer $\mu=n$.
The opposite implication follows Corollary \ref{cor:pdes}.
That is to say,
equation (\ref{eq:main1}) for a particular positive integer $n$
implies equation (\ref{eq:main0}) for $\mu\in[n-1,n]$, and thus
equation (\ref{eq:main0}) for $\mu\ge0$ follows
equation (\ref{eq:main1}) for all positive integers $n$.

Next, equation (\ref{eq:difn}) indicates that
\begin{equation}\label{eq:dxnr}
R_{(n)}(x)
=\frac1{x^{n+1}}\Bigl(x^2\!\frac\rmd{\rmd x}\Bigr)^n\bigl[xR(x)\bigr]
=(-1)^nw^{n+1}\frac{\rmd^n\mathcal R(w)}{\rmd w^n}\biggr\rvert_{w=x^{-1}},
\end{equation}
where
\begin{equation}\label{eq:rlap}
\mathcal R(w)\equiv\frac{R(w^{-1})}w.
\end{equation}
Hence equation (\ref{eq:main1}) is also equivalent to
\begin{equation}\begin{split}
&\Bigl(x^2\!\frac\rmd{\rmd x}\Bigr)^n\bigl[xR(x)\bigr]\ge0
&&(x>0,\,n=0,1,2,\dotsc),
\\
&(-1)^n\frac{\rmd^n\mathcal R(w)}{\rmd w^n}\ge0
&&(w>0,\,n=0,1,2,\dotsc).
\end{split}\end{equation}
The last is equivalent to saying that the function $\mathcal R(w)$
defined in equation (\ref{eq:rlap}) is a \emph{completely monotonic}
(Definition \ref{def:cm}) function of $w$.
The \emph{Bernstein theorem} (Theorem \ref{th:HBW}) then implies that
$\mathcal R(w)$ is representable as
the Laplace transform of a non-negative function.
In other words, there exists a non-negative function $\phi(t)\ge0$
of $t>0$ such that $\mathcal R(w)=\mathcal L_{t\rightarrow w}[\phi(t)]$.
The inverse Laplace transformation may
be found using the \emph{Post--Widder formula} (eq.~\ref{eq:pinv}),
which, thanks to equation (\ref{eq:dxnr}), reduces to
\begin{equation}\label{eq:phit}
\phi(t)\equiv\underset{w\rightarrow t}{\mathcal L^{-1}}[\mathcal R(w)]
=\lim_{n\rightarrow\infty}
\frac1{n!}R_{(n)}\Bigl(\frac tn\Bigr).
\end{equation}
Thus we find another equivalent necessary condition,
\begin{equation}\label{eq:main4}
\lim_{n\rightarrow\infty}
\frac1{n!}\frac{\rmd^n[x^nR(x)]}{\rmd x^n}\biggr\rvert_{x=t/n}\ge0
\qquad(t>0).
\end{equation}
It is obvious that equation (\ref{eq:main1}) implies
equation (\ref{eq:main4}), provided that it converges.
The converse on the other hand
follows the Bernstein theorem and the Post--Widder formula.
However, the conditional equivalence given its convergence
may also be inferred from Corollary \ref{cor:des}. By definition,
equation (\ref{eq:main4}) indicates that there exists
a sufficiently large integer ${}^\exists m>0$ such that $R_{(n)}(x)\ge0$
for all ${}^\forall n\ge m$ and $x>0$. Corollary \ref{cor:des} then
suggests that $R_{(m-1)}(x)\ge0$ for $x>0$, and equation (\ref{eq:main1})
follows subsequent successive arguments with descending subscripts
of $R_{(n)}(x)$.

\subsection{Conditions on the potential part}
\label{sec:potn}

\citet{vHBD11} have proven that given equation (\ref{eq:sep}),
$P^{(k)}(\Psi)\ge0$ for all accessible $\Psi$ and
any non-negative integer $k$ not greater than $\frac32-\beta_0$
%
%
where $\beta_0$ is the limit of the anisotropy parameter at the centre,
is necessary for the df to be non-negative.
We shall show that this generalizes incorporating fractional
derivatives.

If the AD is given as in equation (\ref{eq:sep}),
equation (\ref{eq:dpirn2}) results in
\begin{equation}\label{eq:dpirns}
{{}_{E_0}D_\Psi}^\mu
{{}_0I_x}^{\xi-\frac12}\Bigl(\frac\Nu{x^{\xi-1/2}}\Bigr)
={{}_{E_0}D_\Psi}^\mu P\cdot
{{}_0I_x}^{\xi-\frac12}\Bigl(\frac R{x^{\xi-1/2}}\Bigr)\ge0,
\end{equation}
for $0\le\mu\le\xi$. Since $R(x)\ge0$ is again trivially necessary,
${{}_0I_x}^\lambda(x^{-\lambda}R)>0$
for $x>0$ and any $\lambda\ge0$ unless $R(x)=0$
\emph{almost everywhere} in $x\equiv r^2\in[0,\infty)$
(Lemma \ref{lem:pos}). Ignoring pathological cases,
we conclude that equation (\ref{eq:dpirns}) implies that
\begin{equation}\label{eq:pnc}
0<{{}_0I_x}^\lambda(x^{-\lambda}R)<\infty
\quad\Longrightarrow\
{{}_{E_0}D_\Psi}^\mu P\ge0
\qquad(\mu\le\lambda+\tfrac12).
\end{equation}
With $\lambda=0$, this indicates that ${{}_{E_0}D_\Psi}^\mu P\ge0$
for $\mu\le\frac12$.
For $\lambda>0$ on the other hand, equation (\ref{eq:pnc}) implies that,
if $x^{-\lambda}R(x)\,\rmd x$ is integrable over $x=0$,
then ${{}_{E_0}D_\Psi}^\mu P\ge0$ for $\mu\le\lambda+\frac12$ and
all accessible $\Psi$ is necessary for a non-negative df.
Alternatively, ${{}_{E_0}D_\Psi}^\mu P\ge0$ with a fixed $\mu>\frac12$
is necessary for the df to be non-negative
if there exists ${}^\exists\lambda\ge\mu-\frac12$ such that
${{}_0I_x}^\lambda(x^{-\lambda}R)$ is well-defined.

Equation (\ref{eq:pnc}) is yet inconclusive regarding whether
${{}_{E_0}D_\Psi}^{\frac32-\beta}P\ge0$ is necessary for the phase-space
consistency given $R(x)\sim x^{-\beta}$ with $\beta<1$ as $x\rightarrow0$,
which is in fact necessary as shown follows.
For this, we first note that if $h(t)$ is right-continuous at $t=a$,
\begin{equation}\label{eq:dlim}
\lim_{\epsilon\rightarrow0^+}
\epsilon\!\int_a^{\bar t}\!\frac{h(t)\,\rmd t}{(t-a)^{1-\epsilon}}
=\lim_{t\rightarrow a^+}h(t)=h(a)
\qquad(a<\bar t).
\end{equation}
This applied to the left-hand side of equation (\ref{eq:dpirn2}) results in
\begin{subequations}
\begin{gather}
\lim\nolimits_{\xi\rightarrow(\frac32-\eta)^-}\
\bigl(\tfrac32-\eta-\xi\bigr)\,
{{}_0I_x}^{\xi-\frac12}\Bigl(\frac\Nu{x^{\xi-1/2}}\Bigr)
=\frac{\hat P_\eta(\Psi)}{x^\eta\Gamma(1-\eta)}
\intertext{where $\eta<1$ and}\label{eq:pb}
\hat P_\eta(\Psi)=\lim\nolimits_{x\rightarrow0^+}x^\eta\Nu(\Psi,x).
\end{gather}
Equation (\ref{eq:dpirn2}) then results in the formula,
\begin{equation}\label{eq:nc0}
{{}_{E_0}D_\Psi}^\mu\hat P_\eta(\Psi)
=2^{\frac32-\eta}\pi^\frac32\Gamma(1-\eta)\,
{{}_{E_0}I_\Psi}^{\frac32-\eta-\mu}
\tilde g_\eta(\Psi)\ge0,
\end{equation}
where
\begin{equation}\label{eq:gb}
\tilde g_\eta(E)
=\lim\nolimits_{L^2\rightarrow0^+}L^{2\eta}F(E,L^2).
\end{equation}
\end{subequations}
For $\mu<\frac32-\eta$, this is derived with the limit
$\xi\rightarrow(\frac32-\eta)^-$ while maintaining $\mu<\xi<\frac32-\eta$.
For $\mu=\frac32-\eta$ on the other hand, the same limit is taken
with $\mu=\xi$. Hence, equation (\ref{eq:nc0}) is valid for
$\mu\le\frac32-\eta$ and $\eta<1$,
provided that ${{}_0I_x}^{\xi-\frac12}(x^{\frac12-\xi}\Nu)$
is well-defined for $\xi<\frac32-\eta$ (n.b.,
the integrability of the same for $\xi=\frac32-\eta$ is
actually \emph{not} required for its validity). The non-negativity
of equation (\ref{eq:nc0})
follows the non-negativity of $F(E,L^2)$.
Of particular interests are equation (\ref{eq:nc0}) for $\mu=0$ and
$\frac32-\eta$,
\begin{equation}\begin{split}
\hat P_\eta(\Psi)
&=2^{\frac32-\eta}\pi^\frac32\Gamma(1-\eta)
{{}_{E_0}I_\Psi}^{\frac32-\eta}\tilde g_{\eta}(\Psi);
\\\tilde g_\eta(\Psi)
&=\frac{{{}_{E_0}D_\Psi}^{\frac32-\eta}\hat P_\eta(\Psi)}
{2^{3/2-\eta}\pi^{3/2}\Gamma(1-\eta)},
\end{split}\label{eq:cuinv}\end{equation}
which give explicit formulae for $\hat P_\eta(\Psi)$
and $\tilde g_\eta(\Psi)$ from each other.

For a separable AD given as in equation (\ref{eq:sep}), we have
\begin{equation}
\hat P_\eta(\Psi)=\hat R_\eta P(\Psi)\,;\qquad
\hat R_\eta=\lim\nolimits_{x\rightarrow0^+}x^\eta R(x).
\end{equation}
Therefore, equation (\ref{eq:nc0}) indicates that
\begin{equation}
0<\hat R_\eta<\infty
\quad\Longrightarrow\
{{}_{E_0}D_\Psi}^\mu P\ge0
\qquad(\mu\le\tfrac32-\eta).
\end{equation}
%
That is, if there exists ${}^\exists\eta<1$ such that
$\hat R_\eta$ is a positive finite constant,
then ${{}_{E_0}D_\Psi}^\mu P\ge0$ for ${}^\forall\mu\le\frac32-\eta$.
This encompasses equation (\ref{eq:pnc}), which is seen 
as follows: If $\hat R_\eta$ is non-zero finite for $\eta<1$,
then $R\sim x^{-\eta}$ as $x\rightarrow0$.
Hence ${{}_0I_x}^\lambda(x^{-\lambda}R)$
converges for $\lambda<1-\eta$, and so
if $\mu\le\lambda+\frac12$ and
${{}_0I_x}^\lambda(x^{-\lambda}R)$ is well-defined,
then $\mu<\frac32-\eta$.

For example, with a constant anisotropy system of $R(x)=x^{-\beta}$,
we find that $\hat R_\beta=1$ whilst the convergence condition reduces to
\begin{equation}
{{}_0I_x}^\lambda(x^{-\lambda}R)
=\frac1{\Gamma(\lambda)}\!\int_0^x\!
\frac{(x-s)^{\lambda-1}\,\rmd s}{s^{\lambda+\beta}}
=\frac{\Gamma(1-\beta-\lambda)}{x^\beta\Gamma(1-\beta)}<\infty,
\end{equation}
which converges for $0\le\lambda<1-\beta$.
It follows that equation (\ref{eq:pnc}) indicates that 
${{}_{E_0}D_\Psi}^\mu P\ge0$
for $\mu\le\lambda+\frac12<\frac32-\beta$ is necessary
for the df to be non-negative
whereas equation (\ref{eq:nc0}) suggests the same
for $\mu\le\frac32-\beta$ (and $\beta<1$).

\section{Sufficient conditions for phase-space consistency}
\label{sec:suf}

In the companion paper \citep{vH11}, we derive
the necessary \emph{and} sufficient condition for the df with $E_0=0$
to be non-negative, expressed in terms of
the integro-differential constraints of the AD. This is achieved
by reducing the problem to the Hausdorff moment problem,
according to which the df is non-negative if and only if
the moment sequence of equation (\ref{eq:moms}) is
a \emph{completely monotone sequence}\footnote{A sequence
$(a_0,a_1,a_2,\dotsc)$ is completely monotone if and only if
$(-1)^k\Delta^ka_j\ge0$ for all non-negative integer pairs $k$ and $j$.
Here $\Delta$ is the finite difference operator such that
$\Delta^{k+1}a_j=\Delta^ka_{j+1}-\Delta^ka_j$ and $\Delta^0a_j=a_j$.}.
Since the moment sequence is generated by the AD
using equation (\ref{eq:msad}), this condition
is expressible in terms of finite differences
of integro-differential operations on the AD.

With a separable AD, \citet{vH11} also derive a simple
sufficient (but not necessary) condition composed of
two pieces, each of which only involves the potential
or the radius part separately but not together.
In this paper we derive an alternative sufficient condition
for a separable AD to be resulted from a non-negative df,
which turns out to be equivalent to that of \citet{vH11}.
The derivation here is based on the properties of
completely monotonic functions and the Laplace transform.
In the following, we only consider the case that $E_0=0$
and $L_\mathrm m^2=2r^2\Psi$, that is, the df has a compact support
and $F(E<0,L^2)=0$.

\subsection{Sufficient conditions on a separable augmented density}

Inverting equation (\ref{eq:dist}) for $F(E,L^2)$
is formally equivalent to recovering the two-integral even df,
$F^+(E,J_z^2)$ from the axisymmetric density $\nu[\Psi(R^2,z^2),R^2]$
\citep{HQ93}. One notable inversion formula of this kind is
that of \citet{Ly62} who had utilized the Laplace transform.
This suggests that $\phi(t)$ in equation (\ref{eq:phit})
should be related to $F(E,L^2)$.
In Appendix \ref{sec:inv} we do in fact find that
the df that builds the separable AD of equation (\ref{eq:sep})
with $E_0=0$ is recovered via the inverse Laplace transform given by
\begin{equation}\label{eq:lapdf}
F(E,L^2)=\underset{s\rightarrow E}{\mathcal L^{-1}}
\biggl[\frac{s^\frac32\mathcal P(s)}{(2\pi)^{3/2}}
\phi\Bigl(\frac{sL^2}2\Bigr)\biggr].
\end{equation}
where $\mathcal P(s)\equiv\mathcal L_{\Psi\rightarrow s}[P(\Psi)]$
is the Laplace transformation of $P(\Psi)$ and
$\phi(t)$ is as defined in equation (\ref{eq:phit}).

By the Bernstein theorem, equation (\ref{eq:lapdf})
is non-negative if and only if its Laplace transform
is a completely monotonic function of $s>0$
for all accessible $L^2$. However $\mathcal P(s)$
is already completely monotonic since $P(\Psi)\ge0$.
Thus, that $s^\frac32\phi(sL^2/2)$ is a completely
monotonic function of $s>0$ for any $L^2\ge0$ is in fact
sufficient for the df to be non-negative (Lemma \ref{th:cm}).
Equivalently, since
\begin{equation}
\frac{\rmd^n[t^{\frac32}\phi(t)]}{\rmd t^n}\biggr\rvert_{t=sL^2/2}
=\biggl(\frac{L^2}2\biggr)^{\frac32-n}\!
\frac{\rmd^n}{\rmd s^n}
\biggl[s^{\frac32}\phi\Bigl(\frac{sL^2}2\Bigr)\biggr],
\end{equation}
the condition is equivalent to the complete monotonicity of
$t^\frac32\phi(t)$. Unfortunately, this is too severe to be
physical\footnote{If the Laplace transform of $\phi(t)$ exists,
then $\phi(t)$ cannot diverges faster than $t^{-1}$ as $t\rightarrow0$.
Consequently, $\lim_{t\rightarrow0}t^{3/2}\phi(t)\rightarrow0$
and thus $t^{3/2}\phi(t)$ cannot be completely monotonic
because the limit suggests that $t^{3/2}\phi(t)$ should be negative
or increasing in some interval $t\in(0,t_0)$ where ${}^\exists t_0>0$.},
which is inferred in reference to the constant anisotropy model
given by $R(x)=x^{-\beta}$ and $\phi(t)=t^{-\beta}/\Gamma(1-\beta)$.
The condition for this system reduces to
\begin{equation}
\frac{(\beta-\frac32)_n^+}
{\Gamma(1-\beta)}\frac1{t^{\beta+n-3/2}}\ge0
\qquad(t>0,\,n=0,1,2,\dotsc),
\end{equation}
which cannot be satisfied for any constant $\beta<1$.

Nevertheless, the preceding discussion extends to yield
useful sufficient conditions: that is, for any fixed $\lambda$,
the conditions that
\begin{align}
&(-1)^n\frac{\rmd^n[s^\lambda\mathcal P(s)]}{\rmd s^n}\ge0
&&(s>0,\,n=0,1,2,\dotsc),
\label{eq:pcon0}\\\label{eq:rcon}
&(-1)^n\frac{\rmd^n[t^{\frac32-\lambda}\phi(t)]}{\rmd t^n}\ge0
&&(t>0,\,n=0,1,2,\dotsc)
\end{align}
are jointly sufficient to imply equation (\ref{eq:elap})
being completely monotonic and consequently
the df in equation (\ref{eq:lapdf}) being non-negative.
With increasing $\lambda$,
the constraint in equation (\ref{eq:pcon0}) tightens whereas
the condition in equation (\ref{eq:rcon}) becomes strictly weaker.
In other words, with a larger $\lambda$,
the smaller subset of functions $P(\Psi)$ will lead to
$s^\lambda\mathcal P(s)$ being completely monotonic. At the same time
if $\phi(t)$ satisfies equation (\ref{eq:rcon}) for a fixed
$\lambda=\lambda_0$, the same condition for any larger
$\lambda\ge\lambda_0$ automatically holds.
Both of these are easily inferred using Corollary \ref{cor:cmp}.

\subsubsection{the condition on $R(x)$ equivalent to
equation (\ref{eq:rcon})}

To translate equation (\ref{eq:rcon}) into a direct constraint on $R(x)$,
we first assume the existence of $\phi(t)$, the validity of
equation (\ref{eq:phit}), and its non-negativity, that is,
$\phi(t)\ge0$ for $t>0$, which are all necessary.
Substituting equation (\ref{eq:phit}) into equation (\ref{eq:rcon})
then results in
\begin{equation}\begin{split}
(-1)^n\frac{\rmd^n[t^{\frac32-\lambda}\phi(t)]}{\rmd t^n}
&=\lim_{k\rightarrow\infty}\frac{(-1)^n}{k!}\frac{\rmd^n}{\rmd t^n}
\biggl[t^{\frac32-\lambda}R_{(k)}\Bigl(\frac tk\Bigr)\biggr]
\\&=\lim_{k\rightarrow\infty}\frac{(-1)^n}{k!k^{n+\lambda-3/2}}
\frac{\rmd^n[x^{\frac32-\lambda}R_{(k)}(x)]}{\rmd x^n}
\biggr\rvert_{x=t/k}.
\end{split}\end{equation}
Provided that this converges, equation (\ref{eq:rcon}) is
equivalent to insisting that there exists an integer
${}^\exists m>0$ such that, for all integers ${}^\forall k\ge m$
\begin{equation}\label{eq:rcon0}
(-1)^n\frac{\rmd^n}{\rmd x^n}
\biggl\lbrace x^{\frac32-\lambda}\frac{\rmd^k[x^kR(x)]}{\rmd x^k}
\biggr\rbrace\ge0
\qquad(x>0,\,n=0,1,2,\dotsc).
\end{equation}
In other words, the complete monotonicity of
$x^{\frac32-\lambda}R_{(k)}(x)$ for \emph{all sufficiently large}
integers $k$ is equivalent to equation (\ref{eq:rcon}), that is,
the complete monotonicity of $t^{\frac32-\lambda}\phi(t)$.
In fact, equation (\ref{eq:rcon}) is equivalent to
equation (\ref{eq:rcon0}) for not only all sufficiently
large integers but also all non-negative integers $k$,
which follows successive applications of Theorem \ref{th:cmdes}
with descending subscripts $k$ (the opposite implication is trivial).
Note that the condition as stated in this last form, that is,
equation (\ref{eq:rcon0}) for all non-negative integers $k$,
is the same as noted by \citet{vH11}.

\subsubsection{the condition on $P(\Psi)$ equivalent to
equation (\ref{eq:pcon0})}

Explicit constraints on $P(\Psi)$
resulting from equation (\ref{eq:pcon0}) is expressible
by means of fractional calculus. First,
equations (\ref{eq:lapab}) and (\ref{eq:lapderiv}) indicate that
(n.b., ${{}_0I_\Psi}^{1-\delta}P(0)=0$ from Corollary \ref{eq:if0})
\begin{equation}\label{eq:slp}\begin{split}
s^\lambda\mathcal P(s)
&=s^{\mu+1-(1-\delta)}\underset{\Psi\rightarrow s}{\mathcal L}[P(\Psi)]
=s^{\mu+1}\underset{\Psi\rightarrow s}{\mathcal L}
\bigl[{{}_0I_\Psi}^{1-\delta}P(\Psi)\bigr]
\\&=\underset{\Psi\rightarrow s}{\mathcal L}
\bigl[{{}_0D_\Psi}^\lambda P(\Psi)\bigr]
+{\textstyle\sum_{j=1}^\mu s^{j-1}{{}_0D_\Psi}^{\lambda-j}P(0)}
\end{split}\end{equation}
where $\mu=\lfloor{\lambda}\rfloor$ and $\delta=\lambda-\mu$
($0\le\delta<1$) are the integer floor
and the fractional part of $\lambda$.
This suggests that for $\lambda\ge0$, together
\begin{gather}
{{}_0D_\Psi}^\lambda P(\Psi)\ge0
\qquad(\Psi>0),
\label{eq:cons2}\\\label{eq:cons1}
{{}_0I_\Psi}^{1-\delta}P(0)=
{{}_0D_\Psi}^\delta P(0)=\dotsb=
{{}_0D_\Psi}^{\lambda-1}P(0)=0
\end{gather}
are sufficient for $s^\lambda\mathcal P(s)$ to be completely monotonic.
Note, provided that $P(\Psi)$ is right-continuous at $\Psi=0$,
that ${{}_0I_\Psi}^{1-\delta}P(0)=0$ (Corollary \ref{eq:if0}),
which is taken as granted henceforth.
%
If $\lambda=p+1$ is a positive integer,
equations (\ref{eq:cons2}) and (\ref{eq:cons1}) reduce to
\begin{equation}\label{eq:pcon}
P^{(p+1)}(\Psi)\ge0
\quad\&\quad
P(0)=\dotsb=P^{(p)}(0)=0.
\end{equation}
For $0\le\delta<1$ on the other hand, equation (\ref{eq:cons1})
may also be replaced with the same boundary condition as in
equation (\ref{eq:pcon}). That is to say,
$P^{(0)}(0)=\dotsb=P^{(n)}(0)=0$ actually
implies ${{}_0D_\Psi}^{n+\delta}P(0)=0$ for $0<\delta<1$
(Lemma \ref{th:p0}), and thus it follows that for $\lambda\ge1$,
%
\begin{equation}\label{eq:cons1a}
P^{(0)}(0)=\dotsb=P^{(\lfloor{\lambda}\rfloor-1)}(0)
\end{equation}
%
also implies equation (\ref{eq:cons1})
(they are identical if $\delta=0$).
Therefore, together equations (\ref{eq:cons2}) and (\ref{eq:cons1a})
also consist in a sufficient condition
for $s^\lambda\mathcal P(s)$ to be completely monotonic
at a fixed $\lambda$. The condition as expressed with
equation (\ref{eq:cons1a}) is also useful because
equation (\ref{eq:frda}) indicates that
equation (\ref{eq:cons2}) is then equivalent to
\begin{equation}
{{}_0D_\Psi}^\lambda P
=\frac1{\Gamma(1-\delta)}\frac{\rmd^{1+\mu-n}}{\rmd\Psi^{1+\mu-n}}\!
\int_0^\Psi\!\frac{P^{(n)}(Q)\,\rmd Q}{(\Psi-Q)^\delta}\ge0
\end{equation}
%
where $n$ is any non-negative integer not greater than $\lambda$.

Again, the joint condition of
equations (\ref{eq:cons2}) and (\ref{eq:cons1a})
becomes strictly stronger as $\lambda$
increases in accordance with the restriction
on the complete monotonicity of $s^\lambda\mathcal P(s)$.
This is seen with equation (\ref{eq:difcomb}) for $0\le\epsilon\le\lambda$
given equation (\ref{eq:cons1}) or (\ref{eq:cons1a}), that is,
%
${{}_0I_\Psi}^\epsilon\bigl({{}_0D_\Psi}^\lambda P\bigr)
={{}_0D_\Psi}^{\lambda-\epsilon}P$.
%
Therefore, ${{}_0D_\Psi}^\lambda P(\Psi)\ge0$ implies
${{}_0D_\Psi}^\xi P(\Psi)\ge0$ for $0\le\xi\le\lambda$.
The similar implications of equation (\ref{eq:cons1a})
with descending $\lambda$ are trivial.

\subsection{Constant anisotropy models}
\label{sec:cona}

Let us consider the constant anisotropy model given with
\begin{subequations}
\begin{equation}\label{eq:cbeta}
R(x)=x^{-\beta},\qquad
\mathcal R(w)=w^{\beta-1},\qquad
R_{(n)}(x)=(1-\beta)_n^+x^{-\beta},
\end{equation}
which satisfies the necessary condition in Sect.~\ref{sec:radn}
if and only if $\beta\le1$ (cf., Lemma \ref{th:cmp}).
The function $\phi(t)$ as defined in equation (\ref{eq:phit})
for $\beta<1$ is found using either
$\mathcal L_{s\rightarrow t}[s^{a-1}]=t^{-a}\Gamma(a)$ with $a>0$
or $\lim_{n\rightarrow\infty}(n!n^z)/(1+z)_{n}^+=\Gamma(1+z)$ so that
\begin{equation}
\phi(t)=\frac1{t^\beta\Gamma(1-\beta)}
\qquad(\beta<1).
\end{equation}
\end{subequations}
For $\beta=1$, formally $\phi(t)$ results in the Dirac delta.
Although this case will not be discussed
explicitly here (see Appendix \ref{app:b1} instead),
the following result actually extends for $\beta\le1$.

Equations (\ref{eq:rcon}) and (\ref{eq:rcon0}) now reduce to
\begin{equation}\begin{split}
(-1)^n&\frac{\rmd^n[t^{\frac32-\lambda}\phi(t)]}{\rmd t^n}=
\frac1{\Gamma(1-\beta)}
\frac{(\beta+\lambda-\frac32)_n^+}{t^{\beta+n+\lambda-3/2}}\ge0\,;
\\
(-1)^n&\frac{\rmd^n[x^{\frac32-\lambda}R_{(k)}(x)]}{\rmd x^n}=
(1-\beta)_{k}^+
\frac{(\beta+\lambda-\frac32)_n^+}{x^{\beta+n+\lambda-3/2}}\ge0.
\end{split}\end{equation}
For $\beta<1$, this is equivalent to $\beta+\lambda\ge\frac32$.
It follows that if $R(x)=x^{-\beta}$ with $\frac12-p\le\beta<1$
where $p$ is a non-negative integer, then $P(\Psi)$ satisfying
equation (\ref{eq:pcon}) is sufficient for the existence of a
non-negative df \citep[cf.,][]{CM10}.
In general for any \emph{real} $\lambda>\frac12$,
if $R(x)=x^{-\beta}$ with $\frac32-\lambda\le\beta<1$,
equations (\ref{eq:cons2}) and (\ref{eq:cons1a})
constitute a sufficient condition for the phase-space consistency.

For a fixed $\beta<1$, this indicates that,
if there exists ${}^\exists\lambda\ge\frac32-\beta$
such that equations (\ref{eq:cons2}) and (\ref{eq:cons1a})
hold for $P(\Psi$), then $\Nu=r^{-2\beta}P$
guarantees the non-negativity of the corresponding df.
Here the existence of such $\lambda$ further implies
${{}_0D}^\xi_\Psi P\ge0$ for $0\le{}^\forall\xi\le{}^\exists\lambda$
whilst Sect.~\ref{sec:potn} suggests that
${{}_0D_\Psi}^\mu P\ge0$ for ${}^\forall\mu\le\frac32-\beta$ is
necessary for the df inverted from $\Nu=r^{-2\beta}P$ to be non-negative.
It follows that, if $\Nu(\Psi,r^2)=r^{-2\beta}P(\Psi)$, then
${{}_0D_\Psi}^{\frac32-\beta}P\ge0$
is the necessary \emph{and} sufficient condition
for the phase-space consistency. In fact, here
$P(\Psi)=\hat P_\beta(\Psi)$ and $F(E,L^2)=\tilde g_\beta(E)L^{-2\beta}$
where $\hat P_\beta(\Psi)$ and $\tilde g_\beta(E)$ are as defined
in equations (\ref{eq:pb}) and (\ref{eq:gb}) with $\eta=\beta$.
Hence equation (\ref{eq:cuinv}) results in the inversion formula
($\beta<1$),
\begin{equation}
F(E,L^2)
=\frac{{{}_0D_E}^{\frac32-\beta}P(E)}
{2^{3/2-\beta}\pi^{3/2}\Gamma(1-\beta)L^{2\beta}}
\quad\Longleftarrow\
\Nu(\Psi,r^2)=\frac{P(\Psi)}{r^{2\beta}}.
\end{equation}
This is just the generalized \citeauthor{Ed16} inversion formula
\citep[e.g.,][]{EA06} for constant anisotropy systems.
That ${{}_0D_\Psi}^{\frac32-\beta}P(\Psi)\ge0$ is
necessary and sufficient for the existence of a non-negative df
is its trivial consequence.

\section{Family of monotonic anisotropy parameters}
\label{sec:rfun}

Consider the anisotropy parameter \citep{BvH07},
\begin{subequations}
\begin{equation}
\beta(r)
=\frac{\beta_1r_\mathrm a^{2s}+\beta_2r^{2s}}
{r_\mathrm a^{2s}+r^{2s}}
\qquad(s>0,\,r_\mathrm a>0).
\end{equation}
If the spherical system is characterized by a separable AD
as in equation (\ref{eq:sep}), this follows the radial function
(cf., eq.~\ref{eq:beta})
\begin{gather}\label{eq:rfun}
R(x)=x^{-\beta_1}(1+x^s)^{-\zeta}
\qquad\text{where $s\zeta=\beta_2-\beta_1$};
\\\mathcal R(w)=w^{-1}R(w^{-1})
=w^{\beta_1-1}(1+w^{-s})^{-\zeta}
=w^{\beta_2-1}(1+w^s)^{-\zeta}.
\nonumber\end{gather}
\end{subequations}
Hereafter we set $r_\mathrm a=1$ (i.e., $x=r^2/r_\mathrm a^2$), but
this has no effect on the following discussion whatsoever.

Note $R_{(1)}(x)\ge0$ for $x>0$ restricts $\beta_1,\beta_2\le1$.
In fact,
\begin{theorem}[\citealt{An11}]
$R(x)$ given by equation (\ref{eq:rfun}) with 
$0<s\le1$ and $\beta_1,\beta_2\le1$ satisfies
the necessary condition in Sect.~\ref{sec:radn},
\end{theorem}
which is easily deduced from Corollary \ref{cor:qcm}.
However, the situation for $s>1$ is inconclusive.
On one hand, if $\beta_2=1>\beta_1$, then
$\mathcal R''(w)<0$ for $w^s<(s-1)/(2-\beta_1)$
and so the condition fails for $s>1$.
\citet{An11} on the other hand has found that
the condition is met for all $s>0$
if $\zeta$ is zero or a negative integer. It appears
that for $s>1$, there may exist a \emph{proper} subset of parameter
combinations $\beta_1,\beta_2\le1$ that satisfies the necessary condition
of equation (\ref{eq:main1}), but we have not been able to establish
the concrete criteria.

The necessary condition on the potential part in Sect.~\ref{sec:potn}
on the other hand is straightforward since $R(x)\sim x^{-\beta_1}$
as $x\rightarrow0$. That is,
\begin{theorem}
if the AD is given by equation (\ref{eq:sep}) with
$R(x)$ of equation (\ref{eq:rfun}), the potential part
$P(\Psi)$ must satisfy
\begin{equation}
{{}_{E_0}D_\Psi}^\lambda P(\Psi)\ge0
\qquad\text{for ${}^\forall\lambda\le\tfrac32-\beta_1$}
\end{equation}
in order for the df to be non-negative.
\end{theorem}
Here also note $\beta_1\le1$ and thus ${{}_{E_0}D_\Psi}^\lambda P\ge0$
for any $\lambda\le\frac12$.

\subsection{\boldmath Sufficient conditions for a non-negative df
with $0<s\le1$}
\label{app:p1}

By Theorem \ref{eq:gmllt}, equation (\ref{eq:phit}) results in
\begin{equation}\label{eq:phigml}
\phi(t)=t^{-\beta_1}E_{s,1-\beta_1}^\zeta\!\bigl(-t^s\bigr)
\end{equation}
for $R(x)$ in equation (\ref{eq:rfun}) with $s>0$ and
$\beta_1<1$ (for $\beta_1=1$ see Appendix \ref{app:b1}).
Here $E^\lambda_{p,b}(z)$ is
as defined in equation (\ref{def:gml}).

We consider sufficient conditions to guarantee
the phase-space consistency for a separable AD with
$R(x)$ in equation (\ref{eq:rfun}) with $0<s\le1$ (and $E_0=0$).
In Sect.~\ref{sec:cona}, we have argued that for $\beta_1=\beta_2<1$,
if there exists ${}^\exists\lambda\ge\frac32-\beta_1$
such that ${{}_0D_\Psi}^\lambda P\ge0$ and
$P(0)=\dotsb=P^{(\lfloor{\lambda}\rfloor-1)}(0)=0$, then the df with $E_0=0$
inverted from $\Nu=r^{-2\beta_1}P(\Psi)$ is non-negative everywhere.
This follows from the fact that
$t^{\frac32-\lambda}\phi(t)=t^{\frac32-\lambda-\beta_1}/\Gamma(1-\beta)$
is completely monotonic for $\lambda\ge\frac32-\beta_1$.
As with $\phi(t)$ in equation (\ref{eq:phigml}), if $\zeta>0$,
then $t^{\frac32-\lambda}\phi(t)$ is completely monotonic
for $\lambda\ge\frac32-\beta_1$ (Theorem \ref{th:pos}), and thus
\begin{theorem}
for $E_0=0$ and $R(x)$ given by equation (\ref{eq:rfun})
with $0<s\le1$ and $\beta_1<\beta_2\le1$, if there exists
${}^\exists\lambda\ge\frac32-\beta_1$ such that
${{}_0D_\Psi}^\lambda P\ge0$ and
$P(0)=\dotsb=P^{(\lfloor{\lambda}\rfloor-1)}(0)=0$,
then the df inverted from $\Nu=P(\Psi)R(r^2)$ is non-negative.
\end{theorem}
This actually extends to $\beta_1\le\beta_2\le1$
(Sect.~\ref{sec:cona} and Appendix \ref{app:b1}).
Also the $(s,\beta_2)=(1,1)$ case results in the \citeauthor{Cu91}
system and thus this with an integer $\lambda\ge\frac32-\beta_1$
reproduces the sufficient condition of
\citet[eq.~27 or 28 with $m=\lfloor{\frac32-\beta_1}\rfloor$]{CM10}.
Finally if $P(0)=\dotsb=P^{(\lfloor{\frac12-\beta}\rfloor)}(0)=0$,
then ${{}_0D_\Psi}^{\frac32-\beta_1}P\ge0$ is the necessary
\emph{and} sufficient condition for the phase-space consistency
given $E_0=0$ and $R(x)$ with $0<s\le1$ and
$\beta_1\le\beta_2\le1$.

For $\zeta\le0$ on the other hand, thanks to
Theorems \ref{th:nneg} and \ref{th:neg}
(see again Appendix \ref{app:b1} for $\beta_1=1$),
\begin{theorem}
for $E_0=0$ and $R(x)$ given by equation (\ref{eq:rfun})
with $0<s\le1$ and $\beta_2\le\beta_1\le1$, if there exists
${}^\exists\lambda\ge\frac32-\beta_1+sn$ where
$n=\lceil{(\beta_1-\beta_2)/s}\rceil$ is the integer ceiling of
(i.e., the smallest integer that is not less than) $(\beta_1-\beta_2)/s$
such that ${{}_0D_\Psi}^\lambda P\ge0$ and
$P(0)=\dotsb=P^{(\lfloor{\lambda}\rfloor-1)}(0)=0$,
then the df inverted from $\Nu=P(\Psi)R(r^2)$ is non-negative.
\end{theorem}
\begin{theorem}
For $E_0=0$ and $R(x)$ given by equation (\ref{eq:rfun})
with $0<s\le1$, $\beta_2\le\beta_1\le1$, and $\beta_2\le1-s$,
if there exists ${}^\exists\lambda\ge\frac32-\beta_2$ such that
${{}_0D_\Psi}^\lambda P\ge0$ and
$P(0)=\dotsb=P^{(\lfloor{\lambda}\rfloor-1)}(0)=0$,
then the df inverted from $\Nu=P(\Psi)R(r^2)$ is non-negative.
\end{theorem}

\section{Summary}
\label{sec:sum}

The main findings of this paper is summarized are follows:
\begin{itemize}
\item We have argued that a unique augmented density $\Nu(\Psi,r^2)$
(and subsequently the distribution function) is specified
given the potential $\Psi(r)$ and the density profile $\nu(r)$
once the infinite set of the radial velocity moments in every order
(equivalently the complete radial velocity distribution)
as a function of the radius are available \citep[cf.,][]{DM92}.
\item We have also shown that the set of fractional
calculus operations on the augmented density listed in
equation (\ref{eq:msad}) provides with the complete
moment sequence of the distribution function along
$K(E,L^2;\Psi,r^2)=0$ as shown in
equation (\ref{eq:moms}). We infer from this that
the augmented density that ensures the non-negativity
of the distribution function may be deduced by analogy
to the classical moment problem in probability theory \citep{vH11}.
\item This introduces the set of necessary conditions on the augmented
density for the non-negativity of the distribution function.
If the augmented density is multiplicatively separable into
functions of the potential and the radius dependencies like
equation (\ref{eq:sep}), this results in the necessary condition
stated by \citet{An11}, that is, equation (\ref{eq:main1})
for the radius part of the augmented density. We have also
discovered a few equivalent statements of this condition, notably
the complete monotonicity of the function $\mathcal R(w)$
defined in equation (\ref{eq:rlap}) as well as equation (\ref{eq:main4}).
\item The similar argument for the potential part of a separable
augmented density on the other hand recovers the conditions
derived by \citet{vHBD11} and \citet{An11a}, which are further
generalized with fractional calculus to indicate that:
${{}_{E_0}D_\Psi}^\mu P\ge0$ for
all accessible $\Psi$ is necessary if $\mu\le\frac12$ or
there exists ${}^\exists\lambda\ge\mu-\frac12$ such that
${{}_0I_{r^2}}^\lambda[r^{-2\lambda}R(r^2)]$ is well-defined
or ${}^\exists\beta\le\frac32-\mu$ such that
$\lim_{r^2\rightarrow0^+}r^{2\beta}R(r^2)$ is non-zero and finite.
\item The distribution function of
an escapable system with a separable augmented density
may be inverted from the latter utilzing the inverse Laplace transform
as in equation \ref{eq:lapdf}).
The non-negativity of the resulting distribution function is guaranteed
if its Laplace transformation is completely monotonic.
From this we have found that the joint condition at a fixed $\lambda$
composed of equation (\ref{eq:rcon0}) for $R(x)$ with all non-negative
integer pairs $n$ and $k$,
and equations (\ref{eq:cons2}) and (\ref{eq:cons1a}) for $P(\Psi)$
is sufficient to imply the phase-space consistency of the system
corresponding to 
$\Nu(\Psi,r^2)=P(\Psi)R(r^2)$.
\item With $R(x)$ given by equation (\ref{eq:rfun}) with
$0<s\le1$ and $\beta_1,\beta_2\le1$, the condition
${{}_{E_0}D_\Psi}^\lambda P\ge0$ for ${}^\forall\lambda\le\frac32-\beta_1$
is necessary in order for the augmented density $P(\Psi)R(r^2)$
to correspond to a non-negative distribution function.
For an escapable system with the same $R(x)$, if there exists
${}^\exists\lambda\ge\frac32-\min(\beta_1,\beta_2)$
such that equations (\ref{eq:cons2}) and (\ref{eq:cons1a})
hold for $P(\Psi)$, then the augmented density $P(\Psi)R(r^2)$
guarantees the phase-space consistency,
unless $1-p<\beta_2<\beta_1<1$. If $1-p<\beta_2<\beta_1<1$
on the other hand, we at this point
only find a slightly restrictive sufficient condition with
${}^\exists\lambda\ge\frac32-(\beta_1-p)>
\frac32-\beta_2>\frac32-\beta_1>\frac12$
(n.b., $\beta_1-p<1-p<\beta_2<\beta_1<1$).
\end{itemize}

Finally, we briefly consider possible generalizations of
our conditions to inseparable augmented densities. First we note
that it is possible to write down the necessary \emph{and} sufficient
condition for the phase-space consistency of any (i.e., not necessarily
separable) augmented density by means of completely monotone sequences 
as developed by \citet{vH11} although its actual algebraic expression
appears to be rather cumbersome. Secondly, whilst the necessary
conditions discussed in Sect.~\ref{sec:nec} are not directly applicable
for inseparable augmented densities, the idea behind their derivations
is none the less valid in general and straightforward to extend for
arbitrary augmented densities. Lastly, if the augmented density were
to given by a sum of separable components, the joint sufficient
conditions applied for each component are sufficient for
the phase-space consistency of the whole system thanks to the linearity
of the transformation from the df to the AD (however, the similar
argument for the necessary condition is invalid).

\section*{acknowledgments}\small\noindent
In response to the referee's report, the presentation
of the current version (especially in the main body) is substantially
streamlined. For more details, please
refer the publicly available extended version \citep{orig}.
JA appreciate hospitality during his visit to Ghent.
He is supported by the Chinese Academy of Sciences (CAS) Fellowships
for Young International Scientist, grant no.~2009Y2AJ7 and
the National Natural Science Foundation of China (NSFC) Research Fund
for International Young Scientist.

\normalsize
\renewcommand\thedefinition{\thesection\arabic{definition}}
\appendix

\section{Mathematical Preliminary}
\label{sec:math}

\subsection{Fractional calculus}
\label{app:frd}

Although it is not usually a part of typical curricula of
mathematical methods, the concept of fractional calculus,
if not by its name, appears not infrequently in problems
of dynamical systems \citep[e.g.,][]{La81}.
For more backgrounds and details 
see e.g., \citet{SS01} and reference therein.

\begin{definition}
For any non-negative real $\lambda\ge0$,
the \emph{Riemann-Liouville integral} operator is defined to be
\begin{equation}\label{eq:mint}
{{}_aI_x}^\lambda f
\equiv\begin{cases}{\displaystyle
\frac1{\Gamma(\lambda)}\!\int_a^x\!(x-y)^{\lambda-1}f(y)\,\rmd y
}&(\lambda>0)\\f(x)&(\lambda=0)\end{cases},
\end{equation}
where $\Gamma(x)$ is the gamma function.
\end{definition}
%
For $0<\lambda<1$, this is also recognized as the Abel transform with
the classical case corresponding to $\lambda=\frac12$.
Next we define
\begin{definition}
the \emph{fractional derivative} for $\lambda\ge0$ given by
\begin{multline}\label{eq:frd}
{{}_aD_x}^\lambda f
\equiv\frac{\rmd^{\lceil{\lambda}\rceil}}{\rmd x^{\lceil{\lambda}\rceil}}
{{}_aI_x}^{\lceil{\lambda}\rceil-\lambda}f
\\=\begin{cases}{\displaystyle
\frac1{\Gamma(\lceil{\lambda}\rceil-\lambda)}
\frac{\rmd^{\lceil{\lambda}\rceil}}{\rmd x^{\lceil{\lambda}\rceil}}\!
\int^x_a\!\frac{f(y)\,\rmd y}{(x-y)^{\lambda-\lfloor{\lambda}\rfloor}}
}&(\lfloor{\lambda}\rfloor<\lambda<\lceil{\lambda}\rceil)
\\f^{(\lambda)}(x)&(\lambda=\lfloor{\lambda}\rfloor=\lceil{\lambda}\rceil)\end{cases},
\end{multline}
where $\lceil{\lambda}\rceil$ and $\lfloor{\lambda}\rfloor$
are the integer ceiling and floor of $\lambda$, respectively.
\end{definition}
%
The definitions are extended to include a negative index using 
\begin{definition}
for arbitrary real $\lambda$,
\begin{equation}\label{eq:mintn}
{{}_aI_x}^{-\lambda}f={{}_aD_x}^\lambda f
\quad\text{\it and vice versa}.
\end{equation}
\end{definition}
The basic result regarding these operators is the composite rules
\begin{equation}\label{eq:intcomp}\begin{split}
{}_a&{I_x}^\xi\bigl({{}_aI_x}^\lambda f\bigr)
={{}_aI_x}^{\xi+\lambda}f,\\
{}_a&{D_x}^\xi\bigl({{}_aI_x}^\lambda f\bigr)=\begin{cases}
{{}_aI_x}^{\lambda-\xi}f&(\xi\le\lambda)\\
{{}_aD_x}^{\xi-\lambda}f&(\xi\ge\lambda)\end{cases}
\end{split}\end{equation}
for $\lambda,\xi\ge0$, 
provided that all the integrals in their
definitions absolutely converge.
These are shown by direct calculations utilizing the Fubini theorem
and the Euler integral of the first kind for the beta function.
Equations (\ref{eq:intcomp}) are however not
valid for negative indices $\lambda$ or $\xi$ without
modification involving the boundary terms.

For proper results, we first observe for $\xi\ge0$ that
\begin{equation}\label{eq:ibpr}
{{}_aI_x}^{\xi+1}f'(x)={{}_aI_x}^\xi f(x)
-\frac{(x-a)^\xi f(a)}{\Gamma(1+\xi)}.
\end{equation}
For $\xi>0$, this is shown via integration by part
whilst the $\xi=0$ case results from the fundamental
theorem of calculus. Using equations (\ref{eq:intcomp})
and (\ref{eq:ibpr}) (and Corollary \ref{eq:if0}),
we then find that for $\lambda,\xi\ge0$,
\begin{equation}\label{eq:difcomb}\begin{split}
{}_a&{I_x}^\xi\bigl({{}_aD_x}^\lambda f\bigr)
={{}_aD_x}^\lambda\bigl({{}_aI_x}^\xi f\bigr)
-\sum_{k=1}^{\flr\lambda}
\frac{(\xi)_k^-{{}_aD_x}^{\lambda-k}f(a)}{\Gamma(1+\xi)}(x-a)^{\xi-k},\\
{}_a&{D_x}^\xi\bigl({{}_aD_x}^\lambda f\bigr)
={{}_aD_x}^{\xi+\lambda}f
-\sum_{k=1}^{\flr\lambda}
\frac{(-1)^{n+k}(\delta)_{n+k}^+}{\Gamma(1-\delta)}
\frac{{{}_aD_x}^{\lambda-k}f(a)}{(x-a)^{k+\xi}}
\end{split}\end{equation}
where $n=\flr\xi$ and $\delta=\xi-\flr\xi$,
assuming that all the integrals in their
definitions absolutely converge. Here
\begin{displaymath}{\textstyle
(a)_n^+\equiv\prod_{j=1}^n(a-1+j)\,;\qquad
(a)_n^-\equiv\prod_{j=1}^n(a+1-j)
}\end{displaymath}
are the \emph{rising} and \emph{falling} sequential products,
which are related to each other via $(-a)_n^-=(-1)^n(a)_n^+$
and $(a)_n^-=(a-n+1)_n^+$. Both 
are also referred to as the Pochhammer symbol:
$(a)_n^+$ follows the analyst's convention
whilst $(a)_n^-$ does the combinatorist's.
%
Equation (\ref{eq:ibpr}) also implies that the fractional derivative
of a positive non-integer order may alternatively be given by
\begin{equation}\label{eq:frda}
{{}_aD_x}^\lambda f
=\frac{\rmd^{\clg\lambda-n}}{\rmd x^{\clg\lambda-n}}
{{}_aI_x}^{\clg\lambda-\lambda}f^{(n)}
+\sum_{k=0}^{n-1}
\frac{(-1)^{\flr\lambda-k}(\delta)_{\flr\lambda-k}^+\,f^{(k)}(a)}
{\Gamma(1-\delta)\,(x-a)^{\lambda-k}},
\end{equation}
where $\delta=\lambda-\flr\lambda$ is the fractional part of
$\lambda$ and $n=0,1,\dotsc,\clg\lambda$.

We formalize a fact,
which is important for our purpose, namely
\addtocounter{definition}3
\begin{lemma}\label{lem:pos}
for $\lambda>0$ and $x>a$, if $f(y)\ge0$ for ${}^\forall y\in[a,x]$, then
${{}_aI_x}^\lambda f(x)>0$,
unless $f=0$ \emph{almost everywhere} in $[a,x]$, that is,
provided that the support of $f$ in $(a,x)$ has non-zero measure.
\end{lemma}
This is trivial by the definition of ${{}_aI_x}^\lambda$. Next we note
\begin{lemma}
for a finite $a$,
\begin{equation}
{{}_aI_x}^\lambda f(x)\sim
\frac{f(a)}{\Gamma(\lambda+1)}\,(x-a)^\lambda
\qquad\text{as $x\rightarrow a^+$}
\end{equation}
which is valid for $\lambda\ge0$ if $f(x)$ is right-continuous at $x=a$
or for $\lambda\ge-1$ if $f(x)$ is right-differentiable at $x=a$.
\end{lemma}
This immediately implies that
\begin{corollary}\label{eq:if0}
if $f(x)$ is right-continuous at $x=a$ ($a\ne\pm\infty$)
and $f(a)$ is finite, then
${{}_aI_x}^\lambda f(a)=0$ for $\lambda>0$.
\end{corollary}


Next we examine the behaviour of fractional calculus operators
under the Laplace transform.
The basic result is
for $\lambda\ge0$,
\begin{equation}\label{eq:lapab}
s^{-\lambda}\underset{x\rightarrow s}{\mathcal L}[f(x)]
=\underset{x\rightarrow s}{\mathcal L}\bigl[{{}_0I_x}^\lambda f(x)\bigr].
\end{equation}
This is shown through direct calculations utilizing
the Fubini theorem and the Euler integral of the second kind
for the gamma function.
The Laplace transform of fractional derivatives is then
found by combining equation (\ref{eq:lapab}) with
\begin{equation}\label{eq:lapderiv}
s^{n+1}\underset{x\rightarrow s}{\mathcal L}[f(x)]
=\underset{x\rightarrow s}{\mathcal L}[f^{(n+1)}(x)]
+{\textstyle\sum_{j=0}^ns^jf^{(n-j)}(0)},
\end{equation}
which is valid given that the Laplace transform converges.
Note 
equation (\ref{eq:lapderiv}) is proven for $n=0$ via integration by part
and the induction completes its proof for any non-negative integer $n$.

\subsection{Post--Widder formula \& completely monotonic functions}

\addtocounter{definition}1
\begin{theorem}[\emph{Post}--Widder]
If $\phi(t)$ is continuous for $t\ge0$ and there exist
reals ${}^\exists A>0$ and ${}^\exists b$ such that
$\rme^{-bt}\lvert{\phi(t)}\rvert\le A$
for all ${}^\forall t>0$, then the Laplace transform,
$\mathcal L_{t\rightarrow x}[\phi(t)]
\equiv\int_0^\infty\!\rmd t\,\rme^{-xt}\phi(t)$
converges and is infinitely differentiable in $x>b$.
Moreover, $\phi(t)$ may be inverted from its Laplace transformation
$f(x)=\mathcal L_{t\rightarrow x}[\phi(t)]$ via
the differential inversion formula \citep{Po30,Wi41},
\begin{equation}\label{eq:pinv}
\phi(t)
=\lim_{n\rightarrow\infty}
\frac{(-1)^n}{n!}\,\Bigl(\frac nt\Bigr)^{n+1}
f^{(n)}\Bigl(\frac nt\Bigr)
\qquad(t>0).
\end{equation}
This formula is usually named after Emil Leon Post (1897-1954)
or together with David Vernon Widder (1898-1990).
The proof may be found in a standard text on the Laplace transform.
\end{theorem}

\begin{definition}\label{def:cm}
A smooth function $f(t)$ of $t>0$ is said to be
\emph{completely monotonic} (cm henceforth) if and only if
\begin{equation}
(-1)^nf^{(n)}(t)\ge0
\qquad(t>0,\,n=0,1,2,\dotsc).
\end{equation}
\end{definition}
The archetypal example of cm functions is $f(t)=\rme^{-t}$.
Other elementary examples of cm functions include:
\begin{lemma}\label{th:cmp}
$f(t)=\ln(1+t^{-1})$ is a cm function of $t>0$ whilst
$f(t)=t^{-\delta}$ for $t>0$ is cm if and only if $\delta\ge0$.
\end{lemma}
{\it proof.}
This is shown via direct calculations. That is, for $n\ge0$
\begin{gather}
\frac{\rmd^{n+1}\ln(1+t^{-1})}{\rmd t^{n+1}}
=(-1)^{n+1}n!\,\biggl[\frac1{t^{n+1}}-\frac1{(1+t)^{n+1}}\biggr]
\,;\\
\frac{\rmd^nt^{-\delta}}{\rmd t^n}=(-\delta)_n^-t^{-\delta-n}
=(-1)^n\frac{(\delta)_n^+}{t^{n+\delta}}
\qquad\square.
\end{gather}

Some basic properties of cm functions are:
\renewcommand\theenumi{\it\arabic{enumi}.}
\begin{lemma}\label{th:cm}
Let $f(t)$ and $g(t)$ be cm functions of $t>0$. Then
\begin{enumerate}
\item $(-1)^nf^{(n)}(t)$ for any non-negative integer $n$ is cm.
\item If $F(t)\ge0$ in $(0,\infty)$ and $f(t)=-F'(t)$, then $F(t)$ is cm.
\item $\int_t^\infty\!f(s)\,\rmd s$ is cm, provided that it converges.
\item $af(t)+bg(t)$ is cm where $a$ and $b$ are non-negative constants.
\item $f(t)\cdot g(t)$ is cm.
\item If $F(t)>0$ in $(0,\infty)$ and $f(t)=F'(t)$,
then $(g\circ F)(t)$ is cm.
\item $\exp[f(t)]$ is cm.
\end{enumerate}
\end{lemma}
Here {\it1--4} are trivial whilst {\it5} follows direct calculations
using the Leibniz rule. The last two may be shown by means of
the Fa\`a di Bruno formula, that is,
\begin{equation}\label{eq:fdb}{\textstyle
(g\circ F)^{(n)}(t)
=\sum_{k=0}^ng^{(k)}\bigl[F(t)\bigr]\cdot
B_{n,k}\bigl[f(t),f'(t),\dotsc,f^{(n-k)}(t)\bigr].
}\end{equation}
Here $F'(t)=f(t)$ and $B_{n,k}$ is the Bell polynomial,
\begin{equation}
B_{n,k}(x_0,\dots,x_{n-k})
\equiv\sideset{}{'}\sum_{(j_0,j_1,\dotsc)}
\frac{n!}{j_0!j_1!\dotsm}
\left(\frac{x_0}{1!}\right)^{j_0}
\left(\frac{x_1}{2!}\right)^{j_1}\dotsm.
\end{equation}
where the summation is over all sequences $(j_0,j_1,\dotsc)$
of non-negative integers constrained such that
\begin{equation}{\textstyle
\sum_{m=0}j_m=k\,;\qquad\sum_{m=0}\,(m+1)j_m=n.
}\end{equation}
Note then $\sum_{m=0}mj_m=n-k$ and thus $j_m=0$ for ${}^\forall m>n-k$
(n.b., if otherwise, $j_m\ge1$ for ${}^\exists m>n-k$ and so
$\sum_{m=0}mj_m>n-k$, which is contradictory).
Next, $n-k-\sum_{m=0}j_{2m+1}=2\sum_{m=0}m(j_{2m}+j_{2m+1})$ is even.
This implies that if $f$ is cm, the parity of $B_{n,k}$
in equation (\ref{eq:fdb}) is $(-1)^{n-k}$. Hence,
given that $g$ is also cm, the parity of every term
of equation (\ref{eq:fdb}) is $(-1)^n$, which proves {\it6}.
Equation (\ref{eq:fdb}) also indicates that
\begin{equation}\label{eq:expf}
\frac{\rmd^n\exp[f(t)]}{\rmd t^n}
=\exp[f(t)]\cdot
B_n\bigl[f'(t),f''(t),\dotsc,f^{(n-k+1)}(t)\bigr]
\end{equation}
where $B_n$ is the $n$-th complete Bell polynomial,
\begin{equation}{\textstyle
B_n(x_1,\dots,x_n)\equiv
\sum_{k=1}^nB_{n,k}(x_0,\dots,x_{n-k}).
}\end{equation}
Note $n-\sum_{m=0}j_{2m}=2\sum_{m=0}m(j_{2m-1}+j_{2m})$ is even.
Hence if $f$ is cm, the parity of $B_n$ in equation (\ref{eq:expf})
is $(-1)^n$ and so follows {\it7}.

\begin{corollary}\label{cor:cmp}
Let $g(t)$ be cm, then both
$t^{-\delta}g(t)$ with $\delta\ge0$ and
$g(t^p)$ with $0<p\le1$ are cm.
\smallskip\\{\it proof.}
The first is obvious thanks to Lemmas \ref{th:cmp} and \ref{th:cm}-{\it5}.
The last follows Lemma \ref{th:cm}-{\it6} with $F(t)=t^p$ since
$F'=pt^{p-1}$ for $0<p\le1$ is cm. {\sc q.e.d.}
\end{corollary}

\begin{corollary}\label{cor:qcm}
For $0<p\le1$ and $a,b\ge0$, these are cm:
\begin{equation}
f(t)=t^{-a}(1+t^p)^{-b}\,;\qquad f(t)=t^{-a}(1+t^{-p})^b.
\end{equation}
{\it proof.}
Let $F(t)=c+t^p$. Then $F'=pt^{p-1}$ is cm for $0<p\le 1$.
Hence first $(g\circ F)(t)=(1+t^p)^{-b}$ with $c=1$ and $g(w)=w^{-b}$
for $0<p\le 1$ and $b\ge0$ is cm.
Next, with $c=0$ and $g(w)=b\ln(1+w^{-1})$,
we find that $(g\circ F)(t)=b\ln(1+t^{-p})$ is cm
for $0<p\le1$ and $b\ge0$,
and so is $(1+t^{-p})^b=\exp[b\ln(1+t^{-p})]$.
The final conclusion follows Corollary \ref{cor:cmp}.
{\sc q.e.d.}
\end{corollary}

The fundamental result characterizing cm functions \citep{HBW,Wi41}
is due to
{\!\!\!\!\!\!\cyr Serg\'e\u{i} Nat\'anovich Bernsht\'e\u{i}n}
(Sergei Natanovich Bernstein; 1880-1968),
\begin{theorem}[Hausdorff--\emph{Bernstein}--Widder]\label{th:HBW}\hfill\\
A smooth function $f(x)$ of $x>0$ is completely monotonic
if and only if $f(x)=\int_0^\infty\rme^{-xt}\,\rmd\mu(t)$
where $\mu(t)$ is the Borel measure on $[0,\infty)$,
that is, there exists a non-negative \emph{distribution}
$\phi(t)\ge0$ of $t>0$ such that
$f(x)=\mathcal L_{t\rightarrow x}[\phi(t)]$.
\end{theorem}
The `if'-part is elementary.
Although the complete proof of the `only if'-part
is beyond our scope, the partial proof follows the Post--Widder formula.
That is, if the inverse Laplace transform
$\phi(t)$ 
of a cm function $f(x)$ is well-defined,
then equation (\ref{eq:pinv}), provided that it
converges, indicates that $\phi(t)$
must be non-negative.

\subsection{Generalized Mittag-Leffler function}
\label{app:gml}

Let us consider a particular generalized hypergeometric function
\addtocounter{definition}3
\begin{definition}
\begin{equation}\label{def:gml}
E^\lambda_{p,b}(z)\equiv
\sum_{k=0}^\infty\frac{(\lambda)_k^+}{\Gamma(pk+b)}\frac{z^k}{k!}
\qquad(p>0).
\end{equation}
\end{definition}
This is absolutely convergent for $p>0$ and all $z$, and thus
is an entire function of $z$ with $p>0$. The function defined as such
is the generalization of the Mittag-Leffler function introduced by
\citet[see also \citealt{HMS}]{Pr71} with
$E^1_{p,b}(z)=E_{p,b}(z)$ and $E^1_{p,1}(z)=E_p(z)$.
If $p=1$ on the other hand,
the definition results in
the Kummer confluent hypergeometric function of the first kind, that is,
$E^\lambda_{1,b}(z)={}_1\tilde F_1(\lambda;b;z)
={}_1F_1(\lambda;b;z)/\Gamma(b)$.

Some operational properties of the generalized Mittag-Leffler function
may be derived directly through term-by-term calculations on its
definition. Important for our purpose amongst them are
\begin{gather}\label{eq:gmld}
\frac{\rmd^nE^\lambda_{p,b}({-z})}{\rmd z^n}
=(-1)^n(\lambda)_n^+\,E^{\lambda+n}_{p,b+pn}({-z}),
\\\label{eq:gmli}
(1-\lambda)_n^+{{}_0I_z}^nE^\lambda_{p,b}({-z})
=E^{\lambda-n}_{p,b-pn}({-z})
-\sum_{k=0}^{n-1}\frac{(n-\lambda)_k^-z^k}{k!\Gamma(b-pn+pk)},
\\\label{eq:gmlr}
\frac{\rmd[z^\lambda E^\lambda_{p,b}(-z)]}{\rmd z}
=\lambda z^{\lambda-1}E^{\lambda+1}_{p,b}(-z).
\end{gather}
for a non-negative integer $n$.

Our interest on the generalized Mittag-Leffler function
mostly hinges on the particular Laplace transform, namely
\addtocounter{definition}3
\begin{theorem}\label{eq:gmllt}
for $b,p>0$,
\begin{subequations}
\begin{equation}
\underset{t\rightarrow w}{\mathcal L}
\bigl[t^{b-1}E^\lambda_{p,b}({-t^p})\bigr]
=\frac1{w^b}\biggl(1+\frac1{w^p}\biggr)^{-\lambda}
=\frac1{w^{b-p\lambda}(1+w^p)^\lambda}.
\end{equation}
This is shown by direct term-by-term integrations that result in
\begin{equation}
\int_0^\infty\!\rmd t\,\rme^{-wt}t^{b-1}E^\lambda_{p,b}({-t^p})
=\sum_{k=0}^\infty\frac{(-1)^k(\lambda)_k^+}{k!w^{pk+b}},
\end{equation}
\end{subequations}
and assembling back the binomial expansion of $(1+w^{-p})^{-\lambda}$.
\end{theorem}
%
%
\begin{lemma}\label{lem:gmlp}
If $0<p\le1$, $b>0$, and $b\ge p\lambda$, then
$E^\lambda_{p,b}(-z)\ge0$ is non-negative for all $z>0$.
\smallskip\\{\it proof.}
By Corollary \ref{cor:qcm}, the Laplace transformation
in Theorem \ref{eq:gmllt} is a completely monotonic function
of $w>0$ for $0<p\le1$
either if $b\ge0$ and $\lambda\le0$
or if $b-p\lambda\ge0$ and $\lambda\ge0$.
The Bernstein theorem then indicates that,
if $0<p\le1$, $b>0$, and $b\ge p\lambda$,
then $t^{b-1}E^\lambda_{p,b}({-t^p})\ge0$ for $t>0$ and thus
$E^\lambda_{p,b}({-z})\ge0$ for $z>0$.
{\sc q.e.d.}
\end{lemma}
Given equation (\ref{eq:gmld}), this further indicates that
\begin{theorem}\label{th:pos}
if $0<p\le1$ and $0<p\lambda\le b$, then
$E^\lambda_{p,b}(-z)$ and $E^\lambda_{p,b}(-t^p)$ are
completely monotonic functions of $z>0$ and $t>0$.
\end{theorem}
%
For $\lambda=-\xi\le0$ on the other hand, we find:
\begin{theorem}\label{th:nneg}
If $0<p\le1$, $\xi\ge0$, and $b>0$, then
$z^{-\clg\xi}E^{-\xi}_{p,b}(-z)$ and subsequently
$t^{-p\clg\xi}E^{-\xi}_{p,b}(-t^p)$
are completely monotonic.
\end{theorem}
\begin{theorem}\label{th:neg}
If $0<p\le1$, $\xi\ge0$, $b>0$, and $b\ge p(1-\xi)$, then
$z^{-\xi}E^{-\xi}_{p,b}(-z)$ and $t^{-p\xi}E^{-\xi}_{p,b}(-t^p)$
are completely monotonic.
\end{theorem}
For a non-negative integer $\xi=\clg\xi=\mu$, these are trivial
since $E^{-\mu}_{p,b}(-z)$ then
reduces to a $\mu$-th polynomial of $z$ with all positive coefficients
and subsequently
\begin{equation}
z^{-\mu}E^{-\mu}_{p,b}(-z)
=\sum_{k=0}^\mu\binom\mu k\,\frac{z^{-(\mu-k)}}{\Gamma(b+pk)}.
\end{equation}
Next, equation (\ref{eq:gmld}) for $\lambda=-\xi\le0$ and $n=\clg\xi$
results in 
\begin{subequations}
\begin{equation}
\frac{\rmd^{\clg\xi}E^{-\xi}_{p,b}({-z})}{\rmd z^{\clg\xi}}=
(1-\epsilon)_{\clg\xi}^+\,E^\epsilon_{p,b+p\clg\xi}({-z})
\end{equation}
where $0\le\epsilon=\clg\xi-\xi<1$.
Now it follows equation (\ref{eq:gmli}) that
\begin{equation}
(1-\epsilon)_{\clg\xi}^+{{}_0I_z}^{\clg\xi}
E^\epsilon_{p,b+p\clg\xi}({-z})
=E^{-\xi}_{p,b}({-z})
-\sum_{k=0}^{\clg\xi-1}\binom\xi k\frac{z^k}{\Gamma(b+pk)}.
\end{equation}
\end{subequations}
For $\xi>0$ (n.b., then $\clg\xi\ge1$), this results in
\begin{multline}
z^{-\clg\xi}E^{-\xi}_{p,b}({-z})
=\sum_{k=0}^{\clg\xi-1}\binom\xi k\,\frac{z^{-(\clg\xi-k)}}{\Gamma(b+pk)}.
\\+\frac{(1-\epsilon)_{\clg\xi}^+}{(\clg\xi-1)!}\!
\int_0^1\!\rmd u\,(1-u)^{\clg\xi-1}E^\epsilon_{p,b+p\clg\xi}({-uz}).
\end{multline}
Theorem \ref{th:nneg} (for a non-integer $\xi>0$) follows this since
\begin{equation}
\frac{\rmd^n}{\rmd s^n}\!\int_0^1\!\rmd u\,
(1-u)^kf(su)
=\int_0^1\!\rmd u\,
(1-u)^ku^nf^{(n)}(su),
\end{equation}
and $E^\epsilon_{p,b+p\clg\xi}({-z})$ is cm given
$b+p\clg\xi-p\epsilon=b+p\xi>0$ (Theorem \ref{th:pos}).
Theorem \ref{th:neg} is proven by equation (\ref{eq:gmlr}), that is,
\begin{equation}
-\frac{\rmd[z^{-\xi}E^{-\xi}_{p,b}(-z)]}{\rmd z}
=\frac{\xi E^{1-\xi}_{p,b}({-z})}{z^{\xi+1}}
=\frac{\xi z^{-\clg{\xi-1}}E^{-(\xi-1)}_{p,b}(-z)}{z^{2-\epsilon}},
\end{equation}
which is cm
either if $0<p\le1$, $b>0$, and $\xi\ge1$ (Theorem \ref{th:nneg})
or if $0<p\le1$, $0\le\xi<1$, and $b\ge p(1-\xi)$ (Theorem \ref{th:pos}).

\subsection{Miscellaneous}
\label{sec:prf}

%
\addtocounter{definition}1
\begin{lemma}[\citealt{An11}, theorem~A3]
\begin{equation}\label{eq:difn}
\biggl(x^2\!\frac\rmd{\rmd x}\biggr)^n(xf)
=x^{n+1}\frac{\rmd^n(x^nf)}{\rmd x^n}
\end{equation}
for any non-negative integer $n$ and arbitrary function $f(x)$.
\end{lemma}
This may be proven by induction on $n$. It is also equivalent to
\begin{lemma}[\citealt{An11}, corollary~A4]\label{eq:lem}
\begin{equation}
x^nf_{(n+1)}(x)
=\frac\rmd{\rmd x}\bigl[x^{n+1}f_{(n)}(x)\bigr]
\quad\text{where}\
f_{(n)}(x)\equiv\frac{\rmd^n[x^nf(x)]}{\rmd x^n}.
\end{equation}
\end{lemma}
\begin{corollary}\label{cor:des}
For a non-negative integer $n$,
if $f_{(n+1)}(x)\ge0$ for $x>0$ and $f_{(n)}(0)$ is finite,
then $f_{(n)}(x)\ge0$ for $x>0$,
\end{corollary}
thanks to the fundamental theorem of calculus indicating
\begin{equation}
x^{n+1}f_{(n)}(x)=x^{n+1}f_{(n)}(x)\bigr\rvert_{x=0}
+\int_0^x\!y^nf_{(n+1)}(y)\,\rmd y.
\end{equation}

Lemma \ref{eq:lem} generalizes with fractional calculus.
In particular,
\begin{lemma}
for a non-negative integer $n$ and $0\le\delta<1$,
\begin{equation}\label{eq:dndxnd}\begin{split}
x^{n+1}&{{}_0D_x}^{n+\delta}(x^{n+\delta}f)
={{}_0I_x}^{1-\delta}\bigl[x^{n+\delta}f_{(n+1)}(x)\bigr],
\\x^{n+\delta}&f_{(n+1)}(x)
={{}_0D_x}^{1-\delta}
\bigl[x^{n+1}{{}_0D_x}^{n+\delta}(x^{n+\delta}f)\bigr].
\end{split}\end{equation}
\end{lemma}
which follows
\begin{gather}
{{}_0I_x}^{1-\delta}(x^{n+\delta}f)
=\frac{x^{n+1}}{\Gamma(1-\delta)}\!
\int_0^1\!\frac{t^{n+\delta}f(xt)\,\rmd t}{(1-t)^\delta},
\nonumber\\\begin{split}
{{}_0D_x}^{n+\delta}(x^{n+\delta}f)
&=\frac1{\Gamma(1-\delta)}\!
\int_0^1\!\frac{\rmd t\,t^{n+\delta}}{(1-t)^\delta}
\frac{\rmd^{n+1}[x^{n+1}f(xt)]}{\rmd x^{n+1}}
\\&=\frac1{x^{n+1}\Gamma(1-\delta)}\!
\int_0^x\!\frac{y^{n+\delta}f_{(n+1)}(y)\,\rmd y}{(x-y)^\delta}.
\end{split}
\end{gather}
Note equations (\ref{eq:dndxnd}) for $\delta=0$
reduce to equations (\ref{eq:lem}) and (\ref{cor:des}).
Together Lemmas \ref{lem:pos} and (\ref{eq:dndxnd})
generalize Corollary \ref{cor:des}, 
\begin{corollary}\label{cor:pdes}
for a non-negative integer $n$,
if $f_{(n+1)}(x)\ge0$ for $x>0$,
then ${{}_0D_x}^\mu(x^\mu f)\ge0$ for $x>0$ and $n\le\mu\le n+1$.
\end{corollary}

Corollary \ref{cor:des} may in fact be generalized alternatively, namely,
\begin{theorem}\label{th:cmdes}
for a non-negative integer $n$,
if $x^af_{(n+1)}(x)$ is completely monotonic,
then $x^af_{(n)}(x)$ is also completely monotonic.
\smallskip\\{\it proof.}
Suppose that $x^af_{(n+1)}$ is cm. Then by the Bernstein theorem,
there exists a non-negative function $h(u)\ge0$ of $u>0$ such that
\begin{subequations}
\begin{equation}
x^af_{(n+1)}(x)=\int_0^\infty\!\rmd u\,\rme^{-xu}h(u).
\end{equation}
The complete monotonicity of $x^af_{(n)}$ can then be shown
directly using equation (\ref{cor:des}), which indicates that
\begin{gather}
x^af_{(n)}=x^{a-n-1}\!\int_0^x\!\rmd y\,y^nf_{(n+1)}(y)
=\int_0^1\!\rmd t\,t^{n-a}\!\int_0^\infty\!\rmd u\,\rme^{-xtu}h(u),
\nonumber\\\frac{\rmd^k[x^af_{(n)}]}{\rmd x^k}
=(-1)^k\int_0^1\!\rmd t\,t^{n+k-a}\!
\int_0^\infty\!\rmd u\,\rme^{-xtu}u^kh(u)
\qquad\square.
\end{gather}
\end{subequations}
\end{theorem}

Finally, we also note
\begin{lemma}\label{th:p0}
for a non-negative integer $n$,
if $f^{(n+1)}(a)$ is finite and
$f^{(0)}(a)=\dotsb=f^{(k)}(a)=0$,
then ${{}_aD_x}^{n+\delta}f(a)=0$ for $0\le\delta<1$.
\smallskip\\{\it proof.}
Here we assume $a=0$, but the similar argument holds for any finite $a$
accompanied by a simple translation. First,
\begin{subequations}
\begin{gather}
{{}_0I_x}^{1-\delta}f(x)
=\frac{x^{1-\delta}}{\Gamma(1-\delta)}\!
\int_0^1\!\frac{f(xt)\,\rmd t}{(1-t)^\delta}
\,;\\\label{eq:pd0}
{{}_0D_x}^{n+\delta}f(x)
=\frac1{\Gamma(1-\delta)}\!\int_0^1\!
\frac{\rmd^{n+1}[y^{1-\delta}f(y)]}{\rmd y^{n+1}}
\biggr\rvert_{y=xt}\frac{t^{n+\delta}\,\rmd t}{(1-t)^\delta}.
\end{gather}
Here the latter follows the former because
\begin{equation}
\frac{\rmd^{n+1}[x^{1-\delta}f(xt)]}{\rmd x^{n+1}}
=t^{n+\delta}
\frac{\rmd^{n+1}[y^{1-\delta}f(y)]}{\rmd y^{n+1}}
\biggr\rvert_{y=xt}.
\end{equation}
Finally, given the Leibniz rule,
\begin{multline}
\frac{\rmd^{n+1}[y^{1-\delta}f(y)]}{\rmd y^{n+1}}
=y^{1-\delta}f^{(n+1)}(y)
\\+(1-\delta)\sum_{k=0}^n(-1)^{n-k}\binom{n+1}k\,(\delta)_{n-k}^+
\frac{f^{(k)}(y)}{y^{n+\delta-k}},
\end{multline}
\end{subequations}
which identically vanishes for $y=0$ if
the condition part of Lemma \ref{th:p0} with $a=0$ holds. Here
the conclusion follows as the integrand of equation (\ref{eq:pd0})
with $x=0$ is also zero. {\sc q.e.d.}
\end{lemma}

\section{derivations of equation \lowercase{(\ref{eq:dpr})}}
\label{sec:cal}

First we establish for any $s>-1$ and $\lambda\ge0$ that
\begin{subequations}\label{eq:ints}
\begin{gather}\label{eq:ints1}
{{}_0I_{r^2}}^\lambda\biggl(r^{2s}\!
\iint\limits_T\!\rmd E\,\rmd L^2K^sG\biggr)
=\frac{r^{2(s+\lambda)}}{2^\lambda(s+1)_\lambda^+}\!
\iint\limits_T\!
\frac{K^{s+\lambda}G\,\rmd E\,\rmd L^2}{(\Psi-E)^\lambda};
\\{{}_0I_{r^2}}^\lambda\biggl(\frac1{r^{2\lambda+2}}\!
\iint\limits_T\!\rmd E\,\rmd L^2K^sG\biggr)
=\frac{r^{2\lambda-2}}{(s+1)_\lambda^+}\!
\iint\limits_T\!
\frac{K^{s+\lambda}G\,\rmd E\,\rmd L^2}{L^{2\lambda}};
\\{{}_{E_0}I_\Psi}^\lambda\!
\iint\limits_T\!\rmd E\,\rmd L^2K^sG
=\frac1{2^\lambda(s+1)_\lambda^+}\!
\iint\limits_T\!\rmd E\,\rmd L^2K^{s+\lambda}G,
\end{gather}
\end{subequations}
provided that all integrals converge and
the $\Psi$ and $r^2$ dependencies of an arbitrary integrable
function $G=G(E,L^2)$ are only through $E$ and $L^2$ --
here and henceforth trivial arguments of $G(E,L^2)$
are suppressed for the sake of brevity. In addition,
%
\begin{displaymath}
\frac1{(s+1)_\lambda^+}=
\frac{\Gamma(s+1)}{\Gamma(s+\lambda+1)}=(s)_{-\lambda}^-
\end{displaymath}
is the generalized Pochhammer symbol.
These are demonstrated by direct calculations
utilizing the Fubini theorem that
are identical to that of \citet{An11a} except for
different arguments involved in the Euler integral for
the beta function.
We also find additional properties of the integral transform
in the form of equation (\ref{eq:dist}), namely,
for any $s>-1$ and a non-negative integer $n\ge0$,
\begin{subequations}\label{eq:difsn}
\begin{gather}
\frac{\partial^n}{\partial\Psi^n}\!
\iint_T\!\rmd E\,\rmd L^2K^sG
\nonumber\\\qquad
=\begin{cases}{\displaystyle
2^n(s)_n^-\!\iint_T\!\rmd E\,\rmd L^2K^{s-n}G
}&(n<s+1)\smallskip\\{\displaystyle
2^ss!\!\int_0^{L_\mathrm m^2}\!\rmd L^2
G\Bigl(\Psi-\frac{L^2}{2r^2},L^2\Bigr)
}&(n=s+1)\end{cases},
\\\biggl(r^4\!\frac\partial{\partial r^2}\biggr)^n\!
\iint_T\!\rmd E\,\rmd L^2K^sG
\nonumber\\\qquad
=\begin{cases}{\displaystyle
(s)_n^-\!
\iint_T\!\rmd E\,\rmd L^2K^{s-n}L^{2n}G
}&(n<s+1)\smallskip\\{\displaystyle
\frac{s!}2\!\int_0^{L_\mathrm m^2}\!\rmd L^2
L^{2s+2}G\Bigl(\Psi-\frac{L^2}{2r^2},L^2\Bigr)
}&(n=s+1)\end{cases}.
\label{eq:difsnr}
\end{gather}
\end{subequations}

With $\Nu=m_{0,0}(\Psi,r^2)$ in equation (\ref{eq:dist}),
these then result in
\begin{subequations}
\begin{gather}
\frac{\partial^n}{\partial\Psi^n}\biggl[
{{}_0I_{r^2}}^{\xi-\frac12}\Bigl(\frac{\Nu}{r^{2\xi-1}}\Bigr)\biggr]
\nonumber\\\qquad=\begin{cases}{\displaystyle
\frac{2^{n+1}\pi^\frac32r^{2\xi-3}}{\Gamma(\xi-n)}\!
\iint_T\!\rmd E\,\rmd L^2
\frac{K^{\xi-n-1}}{L^{2\xi-1}}
F(E,L^2)
}&(n<\xi)\smallskip\\{\displaystyle
2^\xi\pi^\frac32r^{2\xi-3}\!
\int_0^{L_\mathrm m^2}\!\frac{\rmd L^2}{L^{2\xi-1}}
F\Bigl(\Psi-\frac{L^2}{2r^2},L^2\Bigr)
}&(n=\xi)\end{cases},
\label{eq:dpirn}\\
\biggl(r^4\!\frac\partial{\partial r^2}\biggr)^n
\Bigl(r^2{{}_{E_0}I_\Psi}^{\xi-\frac12}\Nu\Bigr)
\nonumber\\\qquad=\begin{cases}{\displaystyle
\frac{2^{\frac32-\xi}\pi^\frac32}{\Gamma(\xi-n)}\!
\iint_T\!\rmd E\,\rmd L^2
K^{\xi-n-1}L^{2n}F(E,L^2)
}&(n<\xi)\smallskip\\{\displaystyle
2^{\frac12-\xi}\pi^\frac32\!\int_0^{L_\mathrm m^2}\!\rmd L^2
L^{2\xi}F\Bigl(\Psi-\frac{L^2}{2r^2},L^2\Bigr)
}&(n=\xi)\end{cases}.
\label{eq:dripn}
\end{gather}
\end{subequations}
where $n$ is again a non-negative integer and $\xi\ge\frac12$.

Equation (\ref{eq:dpirn2}) for $\xi\ge\frac12$
is a straightforward generalization
of equation (\ref{eq:dpirn}) from an integer $n$ to a real $\mu\le\xi$,
which is similarly shown through direct calculations using
equations (\ref{eq:ints}) and (\ref{eq:difsn}) assuming all the integrals
converge. Next equation (\ref{eq:dpirn2}) for $\xi=\frac12$ is
identical to equation (\ref{eq:dripn}) with $n=0$ (and $\xi=\frac12-\mu$).
since ${{}_{E_0}I_\Psi}^{\xi-\frac12}\Nu={{}_{E_0}D_\Psi}^{\frac12-\xi}\Nu$.
Hence, it is inferred that equation (\ref{eq:dripn}) is in fact valid for
not only $\xi\ge\frac12$ but also $\xi\ge0$
(n.b., $0\le n\le\xi$ and so if $0\le\xi\le\frac12$, then $n=0$).

A generalization of equation (\ref{eq:dripn})
from an integer $n$ to a real $\mu$ (cf., eq.~\ref{eq:difn})
and the extension of equation (\ref{eq:dpirn2}) to $\xi\ge0$ are
possible although demonstrating them through
direct calculations is comparatively nontrivial.
Instead, we follow an indirect route to derive the generalization
of equation (\ref{eq:dripn}). First, equation (\ref{eq:dripn})
with $(n,\xi)=(0,\mu)$ and equation (\ref{eq:ints1})
with $G=F$ and $(s,\lambda)=(\mu-1,1-\delta)$ where $\delta=\mu-\flr\mu$
together indicate that
\begin{equation}
{{}_0I_{r^2}}^{1-\delta}\Bigl(
r^{2\mu}{{}_{E_0}I_\Psi}^{\mu-\frac12}\Nu\Bigr)
=\frac{\pi^\frac32r^{2\flr\mu}}
{2^{\flr\mu-\frac12}\flr\mu!}\!
\iint\limits_T\!\rmd E\,\rmd L^2
\frac{K^{\flr\mu}F(E,L^2)}
{(\Psi-E)^{1-\delta}}
\end{equation}
for $\mu>0$ and $0<\delta<1$.
Applying $[r^4(\partial/\partial r^2)]^{\flr\mu+1}$
on this after dividing by $r^{2\flr\mu}$ (eq.~\ref{eq:difsnr})
and using equation (\ref{eq:difn}), we find that
\begin{multline}\label{eq:lmom0}
{{}_0D_{r^2}}^\mu\Bigl(
r^{2\mu}{{}_{E_0}I_\Psi}^{\mu-\frac12}\Nu\Bigr)
=\frac{\pi^\frac32}{2^{\mu-\frac12}r^{2\mu+2}}\!
\int_0^{L_\mathrm m^2}\!\rmd L^2L^{2\mu}
F\Bigl(\Psi-\frac{L^2}{2r^2},L^2\Bigr)
\\=(2\pi)^\frac32\!
\int_{E_0}^\Psi\!\rmd E\,(\Psi-E)^\mu
F\bigl[E,2r^2(\Psi-E)\bigr],
\end{multline}
which is the $\xi=\mu$ case of equation (\ref{eq:dripn2}).
Note, thanks to equation (\ref{eq:difn}),
this is consistent with the case $n=\xi$ of
equation (\ref{eq:dripn}).
Thus, equation (\ref{eq:lmom0}) is actually valid
for any $\mu\ge0$ including integer values.
Finally, let us apply ${{}_{E_0}I_\Psi}^{\xi-\mu}$
to equation (\ref{eq:lmom0}). It then follows the Fubini theorem that
for $0\le\mu<\xi$
\begin{multline}\label{eq:lmom}
{{}_0D_{r^2}}^\mu\Bigl(
r^{2\mu}{{}_{E_0}I_\Psi}^{\xi-\frac12}\Nu\Bigr)
\\=\frac{(2\pi)^\frac32}{2^\xi r^{2\mu+2}\Gamma(\xi-\mu)}\!
\iint\limits_T\!\rmd E\,\rmd L^2
K^{\xi-\mu-1}L^{2\mu}F(E,L^2),
\end{multline}
which recovers the remaining part $(\xi>\mu)$ of equation (\ref{eq:dripn2}).
Equations (\ref{eq:lmom0}) and (\ref{eq:lmom}) together
(i.e., eq.~\ref{eq:dripn2}) constitute the generalization of
equation (\ref{eq:dripn}) from an integer $n$ to a real $\mu$,
which is valid for any pair $(\mu,\xi)$ with $0\le\mu\le\xi$.

Lastly, note that the indices transform
$(\mu,\xi)\rightarrow(\frac12-\xi,\frac12-\mu)$ sends
equation (\ref{eq:dpirn2}) to (\ref{eq:dripn2}) and
and vice versa. Therefore equation (\ref{eq:dripn2})
with $0\le\mu\le\xi\le\frac12$ here implies that
equation (\ref{eq:dpirn2}) is also valid for any $\mu$ and $\xi$
with $0\le\mu\le\xi\le\frac12$, too.
%
%
%

\section{Derivation of Equation \lowercase{(\ref{eq:lapdf})}}
\label{sec:inv}

We first apply the Laplace transform on $\Psi$ to
equation (\ref{eq:dist}),
\begin{multline}
\underset{\Psi\rightarrow s}{\mathcal L}\bigl[\Nu(\Psi,r^2)\bigr]
=\int_0^\infty\!\rmd\Psi\,\rme^{-s\Psi}\Nu(\Psi,r^2)
\\=\frac{2\pi}{r^2}\!
\iint_{E\ge0,L^2\ge0}\!\rmd E\,\rmd L^2
F(E,L^2)\!\int_0^\infty\!\rmd\Psi\,\rme^{-s\Psi}
\frac{\Theta(K)}{\sqrt{\lvert{K}\rvert}}.
\end{multline}
The inner integral in the last line reduces to
\begin{equation}
\int_0^\infty\!\rmd\Psi\,\rme^{-s\Psi}
\frac{\Theta(K)}{\sqrt{\lvert{K}\rvert}}
=\sqrt{\frac\pi{2s}}\,\rme^{-sE}
\exp\biggl\lgroup-\frac{sL^2}{2r^2}\biggr\rgroup,
\end{equation}
and consequently we find that
\begin{equation}
\underset{\Psi\rightarrow s}{\mathcal L}[\Nu]
=\frac{\sqrt2\pi^\frac32}{\sqrt sr^2}\!
\int_0^\infty\!\rmd L^2\exp\biggl\lgroup-\frac{sL^2}{2r^2}\biggr\rgroup\!
\int_0^\infty\!\rmd E\,\rme^{-sE}
F(E,L^2).
\end{equation}
Substituting variables,
$t=\frac12sL^2$ and $w=r^{-2}$, this reduces to
\begin{equation}\label{eq:laps}
\underset{\Psi\rightarrow s}{\mathcal L}\bigl[\Nu(\Psi,w^{-1})\bigr]
=\biggl(\frac{2\pi}s\biggr)^\frac32w
\underset{t\rightarrow w}{\mathcal L}
\biggl[\int_0^\infty\!\rmd E\,\rme^{-sE}
F\Bigl(E,\frac{2t}s\Bigr)\biggr].
\end{equation}

If the AD is separable as in equation (\ref{eq:sep}), then
\begin{equation}\label{eq:lapsep}
w^{-1}\underset{\Psi\rightarrow s}{\mathcal L}\bigl[\Nu(\Psi,w^{-1})\bigr]
=\mathcal R(w)\underset{\Psi\rightarrow s}{\mathcal L}[P(\Psi)]
=\mathcal P(s)\underset{t\rightarrow w}{\mathcal L}[\phi(t)]
\end{equation}
where $\mathcal P(s)\equiv\mathcal L_{\Psi\rightarrow s}[P(\Psi)]$
and $\mathcal R(w)=\mathcal L_{t\rightarrow w}[\phi(t)]$.
Given that the inverse Laplace transformation is unique,
equations (\ref{eq:laps}) and (\ref{eq:lapsep}) together
then imply
\begin{equation}
\mathcal P(s)\,\phi(t)
=\biggl(\frac{2\pi}s\biggr)^\frac32\int_0^\infty\!\rmd E\,\rme^{-sE}
F\Bigl(E,\frac{2t}s\Bigr),
\end{equation}
and reinstating $t=\frac12sL^2$ then leads to
\begin{equation}\label{eq:elap}
\frac{s^\frac32\mathcal P(s)}{(2\pi)^{3/2}}
\phi\Bigl(\frac{sL^2}2\Bigr)
=\int_0^\infty\!\rmd E\,\rme^{-sE}
F(E,L^2)
=\underset{E\rightarrow s}{\mathcal L}\bigl[F(E,L^2)\bigr].
\end{equation}
Equation (\ref{eq:lapdf}) is simply the inversion of this.

\section{\boldmath The $\beta_1=1$ cases}
\label{app:b1}

\subsection{\boldmath The $\beta=1$ constant anisotropy model}

Let us consider the df given by
\begin{equation}
\sqrt2\pi^\frac32F(E,L^2)=
f(E)\deltaup(L^2)
\end{equation}
where $f(E)$ is an arbitrary function of $E$
and $\deltaup(L^2)$ is the Dirac delta.
This df corresponds to the spherical system entirely built
by radial orbits, that is,
the $\beta=1$ constant anisotropy model.
Given that $\mathcal K(L^2=0)=2(\Psi-E)$,
the corresponding AD is found to be
\begin{equation}
\Nu(\Psi,r^2)=\frac1{r^2}\!\sqrt{\frac2\pi}\!
\int_{E_0}^\Psi\!
\frac{f(E)\,\rmd E}{\sqrt{2(\Psi-E)}}
=r^{-2}{{}_{E_0}I_\Psi}^\frac12f(\Psi),
\end{equation}
which is separable as in equation (\ref{eq:sep}) with
%
$P(\Psi)={{}_{E_0}I_\Psi}^\frac12f(\Psi)$
and $R(x)=x^{-1}$.
%
The AD is easily inverted to the df,
%
$f(E)={{}_{E_0}D_E}^\frac12P(E)$,
%
whose non-negativity is also the necessary \emph{and}
sufficient condition for the phase-space consistency.
This is consistent with the results of Sect.~\ref{sec:cona}
applicable for $\beta\le1$ as is $R(x)=x^{-1}$ the natural limit of
the constant anisotropy model in equation (\ref{eq:cbeta}) to $\beta=1$.

We find that
${{}_0I_x}^\lambda x^{-1-\lambda}\rightarrow\infty$,
${{}_0I_x}^{1-\delta}x^{\lambda-1}=x^n\Gamma(\lambda)/n!$, and
${{}_0D_x}^\lambda x^{\lambda-1}=0$ for $\lambda=n+\delta>0$,
%
%
whilst ${{}_0I_x}^0x^{-1}={{}_0D_x}^0x^{-1}=x^{-1}$. Hence,
$R=x^{-1}$ satisfies the necessary condition
in equation (\ref{eq:main1}).
Moreover, equations (\ref{eq:dpirn2}) and (\ref{eq:dripn2})
still hold with non-trivial cases indicating
%
${{}_{E_0}D_\Psi}^\mu P
={{}_{E_0}I_\Psi}^{\frac12-\mu}f(\Psi)$,
%
whose non-negativity for ${}^\forall\mu\le\frac12$ is
the same necessary condition for $P(\Psi)$ discussed
in Sect.~\ref{sec:potn}.

From $R(x)=x^{-1}$, we also find $\mathcal R(w)=1$ and
$\phi(t)=\deltaup(t)$. Although equation (\ref{eq:rcon})
strictly is then trivial as $\deltaup(t)=0$ for $t>0$, this
interpretation of equation (\ref{eq:rcon}) seems improper
considering that the Dirac delta is not differentiable at $t=0$.
Equation (\ref{eq:rcon0}) on the other hand reduces to
$x^{\frac12-\lambda}$ being cm since $R_{(0)}(x)=R(x)=x^{-1}$
and $R_{(n)}(x)=0$ for any positive integer $n$. The sufficient
condition following this, that is, equations (\ref{eq:cons2})
and (\ref{eq:cons1a}) for ${}^\exists\lambda\ge\frac12$ is in fact
a proper one, as is the natural limiting case of the constant
anisotropy model for $\beta=1$. It appears that for $R\sim x^{-1}$
as $x\sim0$ (and $\lim_{w\rightarrow\infty}\mathcal R$ being
nonzero finite), we may consider $\phi(t)\sim t^{-1}$
as $t\sim0$ for the purpose of applying equation (\ref{eq:rcon}).

\subsection{\boldmath Equation (\ref{eq:rfun}) with $\beta_1=1$}

The discussion on necessary conditions (Sect.~\ref{sec:nec})
is valid inclusively for $\beta_1\le1$. That is,
equation (\ref{eq:rfun}) with $\beta_1=1$ still requires to satisfy
equation (\ref{eq:main1}) -- if $0<p\le1$, this is automatically met
-- in order for the df to be non-negative
whereas the potential dependent part is restricted to be
${{}_{\mathcal E_0}D_\Psi}^\frac12P\ge0$ for the phase-space
consistency.

The complication arises however for $\beta_1=1$
in regards to sufficient conditions discussed in Sect.~\ref{app:p1}.
The main difficulty is due to the fact that
$\lim_{x\rightarrow0}xR(x)=\lim_{w\rightarrow\infty}\mathcal R(w)=1$
is non-zero.
Whilst this indicate $\phi\sim t^{-1}$ for $t\sim0$, 
%
%
this behaviour is incompatible with the convergence 
of the Laplace transform. The formal solution follows adopting
$\lim_{a\rightarrow1^-}x^{-a}/\Gamma(1-a)=\deltaup(x)$.
Then, the function $\phi(t)$ in equation (\ref{eq:phigml})
with $\beta_1=1$ is in fact the inverse Laplace
transform of ``$\mathcal R(w)-1$'' whilst the `true' inverse transform
of $\mathcal R(w)$ with $\beta_1=1$ is given by ``$\phi(t)+\delta(t)$''.
For example, since $1/\Gamma(0)=0$, the $k=0$ term in
equation (\ref{def:gml}) for $E^\lambda_{p,0}$
does not contribute. Hence, 
equation (\ref{eq:gmllt}) can in fact be well-defined
for the $b=0$ case too. In particular,
$\mathcal L_{t\rightarrow w}[t^{-1}E^\lambda_{p,0}(-t^p)]
=(1+w^{-p})^{-\lambda}-1$.
%
%
Since $(1+w^{-p})^{-\lambda}\ge1$ for $w>0$ and $\lambda\le0$,
it follows that, if $0<p\le1$ and $\lambda\le0$,
this is also cm and $E^\lambda_{p,0}(-z)\ge0$ for $z>0$.
Given that $\mathcal L_{t\rightarrow w}[\deltaup(t)]=1$, we also find
from this that 
%
$\mathcal L_{t\rightarrow w}[\delta(t)+t^{-1}E^{-\xi}_{p,0}({-t^p})]
=(1+w^{-p})^\xi$.

For the specific discussion concerning sufficient conditions
for the phase-space consistency, consider
%
$P(\Psi)R(r^2)=P(\Psi)R_0(r^2)+r^{-2}P(\Psi)$
%
where $R_0(x)=R(x)-x^{-1}$. From the corresponding df
%
%
with $E_0=0$, it is obvious that the corresponding sufficient condition
is together ${{}_0D_\Psi}^\frac12P\ge0$ and
those derived in Sect.~\ref{sec:suf} with $R_0(x)$.
In addition, Theorems \ref{th:pos}-\ref{th:neg}
actually extend to $b=0$ thanks to
the non-negativity of $E^\lambda_{p,0}(-z)\ge0$. It follows that
Theorems in Sect.~\ref{app:p1} also hold inclusively for $\beta_1=1$.

\label{lastpage}
\end{document}